
%

\input amstex
\input amssym
\def\figure#1{\vbox{
	\centerline{\vbox{$$\matrix #1 \endmatrix$$}}
	\medskip
	\noindent{\caption}}}
\def\eps{\varepsilon}
\def\vareps{\varepsilon}
\ifx\undefined\fitfig\relax\else \fi
%
%
%
%
%
%
%
%
%
%
\newcount\fitdivnum
\newcount\fitdivden
\newcount\fitdivtmp
\def\fitdivdig#1{
	\multiply\fitdivtmp by \fitdivden
	\advance\fitdivnum by -\fitdivtmp
	\multiply\fitdivnum by 10%
	\fitdivtmp=\fitdivnum
	\divide\fitdivtmp by \fitdivden
	\edef#1{#1\number\fitdivtmp}}%
\def\fitfpdiv#1#2#3{
	\fitdivnum=#2%
	\fitdivden=#3%
	\ifnum\fitdivnum<0
		\fitdivnum=-\fitdivnum
		\edef#1{-}%
	\else
		\edef#1{}%
	\fi
	\fitdivtmp=\fitdivnum
	\divide\fitdivtmp by \fitdivden
	\edef#1{#1\number\fitdivtmp.}%
	\fitdivdig{#1}%
	\fitdivdig{#1}%
	\fitdivdig{#1}%
	\fitdivdig{#1}%
	\fitdivdig{#1}%
	\fitdivdig{#1}}%
\def\fitfpcdiv#1#2#3{
	\fitdivnum=#2%
	\fitdivden=#3%
	\ifnum\fitdivnum<0
		\fitdivnum=-\fitdivnum
		\edef#1{-}%
	\else
		\edef#1{}%
	\fi
	\fitdivtmp=\fitdivnum
	\divide\fitdivtmp by \fitdivden
	\edef#1{#1\number\fitdivtmp}%
	\fitdivdig{#1}%
	\fitdivdig{#1}%
	\edef#1{#1.}%
	\fitdivdig{#1}%
	\fitdivdig{#1}%
	\fitdivdig{#1}%
	\fitdivdig{#1}}%
%
%
\newdimen\fitllx
\newdimen\fitlly
\newdimen\fiturx
\newdimen\fitury
\newdimen\fitrllx
\newdimen\fitrlly
\newdimen\fitrurx
\newdimen\fitrury
\newdimen\fitpsheight
\newdimen\fitpswidth
\newdimen\fitheight
\newdimen\fitwidth
\newdimen\fithoffset
\newdimen\fitvoffset
\newdimen\fitunit
\fitunit=1bp
\newcount\fitangle
\def\fithscale{}%
\def\fitvscale{}%
\def\fitdraftcenter{DRAFT}
\newif\iffitclip
\newif\iffitdvips
\newif\iffitdraft
\fitclipfalse
\fitdvipstrue
\fitdraftfalse
\def\fitrotate{%
	\ifnum\fitangle=0 
		\fitrllx=\fitllx \fitrlly=\fitlly
		\fitrurx=\fiturx \fitrury=\fitury
	\else\ifnum\fitangle=90
		\fitrllx=-\fitury \fitrlly=\fitllx
		\fitrurx=-\fitlly \fitrury=\fiturx
	\else\ifnum\fitangle=180
		\fitrllx=-\fiturx \fitrlly=-\fitury
		\fitrurx=-\fitllx \fitrury=-\fitlly
	\else\ifnum\fitangle=270
		\fitrllx=\fitlly \fitrlly=-\fiturx
		\fitrurx=\fitury \fitrury=-\fitllx
	\else
		\message{fit: Only angles of 0, %
			90, 180 and 270 are supported.}%
		\fitrllx=\fitllx \fitrlly=\fitlly
		\fitrurx=\fiturx \fitrury=\fitury
	\fi\fi\fi\fi}%
\def\fitscale{%
	\fitpswidth=\fitrurx
	\advance\fitpswidth by -\fitrllx
	\fitpsheight=\fitrury
	\advance\fitpsheight by -\fitrlly
	\ifnum\fitheight=0
		\ifnum\fitwidth=0
			\edef\fithscale{1.000000}%
			\edef\fitvscale{1.000000}%
		\else
			\fitfpdiv\fithscale\fitwidth\fitpswidth
			\edef\fitvscale{\fithscale}%
		\fi
	\else
		\ifnum\fitwidth=0
			\fitfpdiv\fitvscale\fitheight\fitpsheight
			\edef\fithscale{\fitvscale}%
		\else
			\fitfpdiv\fithscale\fitwidth\fitpswidth
			\fitfpdiv\fitvscale\fitheight\fitpsheight
		\fi
	\fi
	\fitwidth=\fithscale\fitpswidth
	\fitheight=\fitvscale\fitpsheight
	\iffitdvips
		\fithoffset=-\fithscale\fitrllx
		\fitvoffset=-\fitvscale\fitrlly
		\fitfpcdiv\fithscale\fitwidth\fitpswidth
		\fitfpcdiv\fitvscale\fitheight\fitpsheight
	\else
		\fithoffset=-\fitrllx
		\fitvoffset=-\fitrlly
	\fi}
\def\fitbox[#1 #2 #3 #4]{%
	\fitllx=#1
	\fitlly=#2
	\fiturx=#3
	\fitury=#4}
\def\fitbpbox[#1 #2 #3 #4]{%
	\fitllx=#1\fitunit
	\fitlly=#2\fitunit
	\fiturx=#3\fitunit
	\fitury=#4\fitunit}

\def\fitxyt[#1 #2 #3]{%
	\fitwidth=#1
	\fitheight=#2
	\fitangle=#3
	\fitrotate
	\fitscale}
\def\fitspecial#1{{\iffitdraft\relax\else
	\iffitdvips\special{%
		psfile=#1\space
		llx=\the\fitllx\space
		lly=\the\fitlly\space
		urx=\the\fiturx\space
		ury=\the\fitury\space
		\iffitclip clip\space\fi
		angle=\number\fitangle\space
		hscale=\fithscale\space
		vscale=\fitvscale\space
		hoffset=\the\fithoffset\space
		voffset=\the\fitvoffset\space}%
	\else
		\fitpswidth=1bp%
		\fitfpdiv\hoffset\fithoffset\fitpswidth
		\fitfpdiv\voffset\fitvoffset\fitpswidth
		\fitfpdiv\llx\fitllx\fitpswidth
		\fitfpdiv\lly\fitlly\fitpswidth
		\fitfpdiv\urx\fiturx\fitpswidth
		\fitfpdiv\ury\fitury\fitpswidth
		\special{%
			psfile=#1\space
			\iffitclip 
				\llx\space\lly\space\urx\space\ury
				doclip\space
			\fi
			rotate=\number\fitangle\space
			hscale=\fithscale\space
			vscale=\fitvscale\space
			hoffset=\hoffset\space
			voffset=\voffset\space
		}\fi\fi}}%
\def\fitfig#1{\iffitdraft\hbox{%
			\vrule height \fitheight width 1pt
			\vbox to \fitheight {%
				\hrule width \fitwidth height 1pt
				\vfil
				\hbox to \fitwidth{\hfil\fitdraftcenter\hfil}%
				\vfil
				\hrule width \fitwidth height 1pt}%
			\vrule height \fitheight width 1pt}%
	\else
		\vbox to\fitheight{\vfil\hbox 
			to\fitwidth{\fitspecial{#1}\hfil}}
	\fi\fitclipfalse}%
\def\fitgz#1{\fitfig{"`gunzip -c #1"}}%
\def\fitembrace#1{[#1]}
\def\fit#1#2#3{%
 	\expandafter\fitbpbox\fitembrace{#1}%
 	\expandafter\fitxyt\fitembrace{#2}%
	\fitfig{#3}}%

\fitdrafttrue

\magnification=\magstep 1
\hsize=6.25 truein
\hcorrection{.125 truein}
\vsize=8.5 truein
\parindent=20pt
\parskip=9pt
\baselineskip=12pt
\NoBlackBoxes
\footline={\hss\tenrm\folio\hss\paging}
\def\paging{\pageno=1}


\def\Schrodinger{{Schr\"odinger }}

\def\eps{{\epsilon}}
\def\vareps{{\varepsilon}}

\def\del{{\partial}}

\def\L {{\hbox{L}}}

\def\C {{\bold C}}
\def\R {{\bold R}}

\def\I {{\bold I}}

\def\REAL{\,{\hbox{Re}}}

\def\BAR{\overline}

\def\tt{{\bold t}}
\def\xx{{\bold x}}

\def\Xp{x^{+}}
\def\Xm{x^{-}}
\def\Xpm{x^{\pm}}

\def\AA{{\Cal A}}
\def\DD{{\Cal D}}

\def\LL{{\Cal L}}
\def\OO{{\Cal O}}

\def\TT{{\Cal T}}

\def\limt{\lim_{\eps \to 0}}

\def\tr{\,\hbox{tr}}





\hbox{}
\bigskip
\bigskip
\centerline{\bf The Zero-Dispersion Limit for the Odd Flows}
\centerline{\bf in the Focusing Zakharov-Shabat Hierarchy}
\bigskip

\bigskip
\centerline{Nicholas M. Ercolani}
\centerline{\it Department of Mathematics}
\centerline{\it University of Arizona, Tucson, AZ 85721}
\medskip
\centerline{Shan Jin}
\centerline{\it MathSoft Inc.}
\centerline{\it 1700 Westlake Dr., Suite 500, Seattle, WA 98109}
\medskip
\centerline{C. David Levermore}
\centerline{\it Department of Mathematics and Institute for Physical Science 
             and Technology}
\centerline{\it University of Maryland, College Park, MD 20742}
\medskip
\centerline{Warren D. MacEvoy Jr.}
\centerline{\it Department of Mathematics}
\centerline{\it Mesa State College, Grand Junction, CO 81502}
\medskip

\bigskip
\centerline{\bf Abstract}
\medskip

We present a numerical and theoretical study of the zero-dispersion
limit of the focusing Zakharov-Shabat hierarchy, which includes NLS
and mKdV flows as its second and third members.  All the odd flows in
the hierachy are shown to preserve real-valued data.  We establish the
zero-dispersion limit of all the nontrivial conserved densities and
associated fluxes for these odd flows for a large class of real-valued 
initial data which includes all ``single hump'' initial data.
In particular, it is done for the ``focusing'' mKdV flow.  The method is 
based on the Lax-Levermore KdV strategy, but here it is carried out in 
the context of a nonselfadjoint spectral problem.  We find that after an 
algebraic transformation the limiting dynamics of the mKdV equation is 
identical to that of the KdV equation.

\newpage

\bigskip
\bigskip
\centerline{\bf 1. Introduction}
\smallskip

\bigskip
\noindent
{\bf 1.1. Background.}  
The asymptotic analysis of nonlinear wavepackets, extending the 
Fourier-theoretic treatment of the linear case, has developed into a 
prominent area of activity in both Analysis and Applied Mathematics.
The first systematic approach to this problem is due to G. Whitham [W], 
who, in the 1960's, found various mechanisms (variational principles, 
averaged conservation laws) for formally deriving {\it modulation equations} 
which describe the
evolution of parameters in a family of exact travelling wave solutions to 
a nonlinear equation. This is in analogy to the classical geometric
optics approach for linear equations. There the asymptotic evolution
of a linear wavepacket is described in terms of a deformation of parameters 
in the family of plane wave solutions corresponding to the dispersion 
relation of the equation.

It was fortuitous that at the same time that Whitham was formulating his 
theory, the subject of completely integrable partial differential equations 
(PDE's), or soliton theory, became widely known. The latter concerns a class 
of physically significant PDE's with large families of exact solutions, 
including multiphase waves, as opposed to just travelling wave subfamilies. 
This provided a rich and fertile testing ground in which to explore the 
techniques that Whitham had developed. Whitham studied a wide variety of 
these equations himself in the travelling wave setting: Korteweg-deVries (KdV),
nonlinear \Schrodinger (NLS), sine-Gordon (SG), etc. The extension of Whitham 
analysis to the class of multiphase solutions was accomplished by Flaschka, 
Forest and McLaughlin [FFM] initially for KdV. This was still in the setting 
of wavepacket analysis. The first work to place Whitham's theory within the 
framework of an explicit singular limit problem was due to Lax and 
Levermore [LL2] who investigated the zero-dispersion limit of the 
Korteweg-deVries 
(KdV) equation.  There one studies the limit as $\eps\to0$ of the conserved 
densities for the scaled KdV equation
$$
\gather
  \del_t u^\eps - 6 u^\eps \del_x u^\eps
                 + \eps^2 \del_{xxx} u^\eps = 0 \,, 
  \tag 1.1a
\\
   u^\eps(x,0) = u_{in}(x) \,.  
  \tag 1.1b
\endgather
$$
The limit is strong and given by the solution of the Hopf equation,
which is hyperbolic,
$$
\gather
   \del_t u + 6 u \, \del_x u = 0 \,, 
  \tag 1.2a
\\
   u(x,0) = u_{in}(x) \,, 
  \tag 1.2b
\endgather
$$
so long as its solution is classical.  After the breaktime the limit 
is weak due to the development of regularizing small wavelength 
oscillations with an amplitude of order unity; thereafter its 
evolution is no longer governed by the Hopf equation (1.2a).

  In their seminal paper, Gardner, Greene, Kruskal and Miura [GGKM]
showed that the KdV equation is completely integrable using the
inverse scattering transform associated with the selfadjoint
\Schrodinger operator
$$
   \LL_{S} = - \eps^2 \del_{xx} + u^\eps \,.
  \tag 1.3
$$
Lax and Levermore [LL] realized that this inverse transform would enable one
to reduce the asymptotic analysis of nonlinear KdV wavepackets to classical
short wave asymptotics.  They analyzed the limiting behavior of the 
scattering and inverse scattering transform using a WKB analysis 
of (1.3) and a kind of steepest descent argument to obtain a 
characterization of the (weak) limits in terms of the solution of 
a variational problem. The solution of this variational problem was 
then constructed through the solution of a Riemann-Hilbert problem.  
Venakides [V] has analyzed the microstructure of the limiting 
solutions, bridging the gap to the local approach of modulation 
theory developed in [FFM].  These results are surveyed in [LLV].

Extensions of the Lax-Levermore strategy to other integrable PDE's have 
been pursued, perhaps the most notable being a recent analysis [JLM2] of
the {\it defocusing} NLS hierarchy.  In this paper we are going to 
investigate the asymptotic behavior of a part of the hierarchy of commuting 
flows associated to the {\it focusing} nonlinear \Schrodinger (NLS) equation
(1.4). A significant difference between the asymptotic analysis of this last 
equation and that of the KdV or even the defocusing NLS hierarchy is that the
inverse transform of the former is based on a non-selfadjoint scattering 
problem (1.5 +) whereas that of the latter is based on selfadjoint scattering
problems. At a formal level this has the consequence that the Whitham 
modulation equations for focusing NLS can become elliptic and therefore 
ill-posed for general initial data. 

This non-selfadjointness also presents 
an obstacle for knowing how to even begin a Lax-Levermore analysis. The work
in this paper makes a first step towards confronting this obstacle. We do not
try to tackle the focusing NLS equation proper; rather, we expand our view and
look at the entire hierarchy of flows.  By abandoning the even members
of the flow, and the possibilty of complex initial data, we see that
the remaining members are a hierarchy in their own right, and that
their limiting behavior is identified with that of the
Korteweg-deVries hierarchy in the same limit.  Numerical simulation
suggests that this is a sharp result: after the initial shock, there
are $O(1)$ differences in the two flows which go weakly to zero as the
dispersion is taken to zero.  

\bigskip
\noindent
{\bf 1.2. The NLS and mKdV Equations.}   
The one dimensional cubic nonlinear \Schrodinger (NLS) focusing $(+)$ and 
defocusing $(-)$ equations
$$
   i \eps \, \del_t u^\eps 
   + \frac{\eps^2}{2} \del_{xx} u^\eps
   \pm |u^\eps|^2 u^\eps = 0 \,,
  \tag 1.4a
$$
with initial data in the amplitude-phase form
$$
   u_\eps(x,0) 
     = A(x) \, \exp\!\bigg( \frac{i}{\eps} S(x) \bigg) \, ,
  \tag 1.4b
$$
were first solved by Zakharov and Shabat [ZS] for some natural far
field boundary conditions of the initial data $A(x)$ (positive) and
$S(x)$ (real) by using the inverse scattering transform associated
with a Dirac operator (ZS operator) [ZS1, ZS2]:
$$
  \LL = \pmatrix -i \eps \, \del_x  &  \pm i\bar{u}^\eps  \\
                     i u^\eps       &  i \eps \, \del_x   \endpmatrix \,.
  \tag 1.5
$$
This ZS operator is selfadjoint if in the defocusing case, but
nonselfadjoint if in the focusing case.  The parameter $\eps$ is
introduced into the evolution and scattering problems with the
zero-dispersion limit scaling; namely, replacing $\del_x$ and
$\del_t$ by $\eps\del_x$ and $\eps\del_t$.  The choice of phase
scaling in the initial data is the natual balance for obtaining a
non-trival momentum density in the limit.  This scaling, both in the
data and the evolution equations, is completely analogous to the
linear \Schrodinger case of quantum mechanics, where the $\eps$ is
taken as a nondimensional version of Planck's constant, $\hbar$.  In
the linear case, there is no specific scattering operator to contend
with.

   For the nonlinear case, the solution of the initial-value problem
centers around the eigenvalue problem 
$$
   \LL f = \zeta f, \qquad \text{where} \qquad 
       f = \pmatrix f^{(1)} \\ f^{(2)} \endpmatrix \,,
$$
where now both $t$ and $\eps$ play roles as parameters, and are not
explicitly differentiated.  For convenience, we may sometimes represent
the eigenvalue $\zeta$ by its real part $\xi$ and imaginary part $\eta$.

   The NLS equations are the second members, and first interesting
members, of a sequence of flows associated with $\LL$ (1.5) [FNR].
The third members are the modified KdV equations (mKdV),
$$
  \del_t u^\eps \pm \frac{3}{2} u^{\eps\,2} \del_x u^\eps 
  + \frac{\eps^2}{4} \del_{xxx} u^\eps = 0 \,,
  \tag 1.6a
$$
or, more generally, their complex forms
$$
  \del_t u^\eps \pm \frac{3}{2} |u^\eps|^2 \del_x u^\eps 
  + \frac{\eps^2}{4} \del_{xxx} u^\eps = 0 \,,
  \tag 1.6b
$$
As members of the focusing $(+)$ and defocusing $(-)$ NLS hierarchies,
they share the same respective ZS operator and the same respective
conserved quantities.

   There are in fact an infinite number of conserved densities for the
NLS hierarchy. The first two of them are mass density $\rho^\eps$ and
momentum density $\mu^\eps$ defined by
$$
  \rho^\eps = |u^\eps|^2 \,, \qquad 
  \mu^\eps  = -i \frac{\eps}{2} 
              \big( \bar{u}^\eps \del_x u^\eps 
                    - u^\eps \del_x \bar{u}^\eps \big) \,.
  \tag 1.7
$$
The corresponding conservation laws of the focusing $(-)$ and
defocusing $(+)$ NLS equations are
$$
\aligned
  \del_t \rho^\eps + \del_x \, \mu^\eps & = 0 \,, 
\\
  \del_t \mu^\eps 
  + \del_x \bigg( \frac{\mu^{\eps\,2}}{\rho^\eps} 
                  \mp \frac{1}{2} \rho^{\eps\,2} \bigg) 
  & = \frac{\eps^2}{4} 
      \del_x \big[ \rho^\eps \del_{xx} \log(\rho^\eps) \big] \,;
\endaligned
  \tag 1.8
$$
while those of the focusing $(+)$ and defocusing $(-)$ mKdV equations
are
$$
\aligned
  \del_t \rho^\eps 
  + \del_x \bigg( \pm \frac{3}{4} \rho^{\eps\,2} 
                  - \frac{3}{4} \dfrac{\mu^{\eps\,2}}
                                      {\rho^\eps} \bigg) 
  & = \frac{\eps^2}{4} 
      \del_x \bigg[ \frac{3}{4} \frac{(\del_x \rho^\eps)^2}{\rho^\eps}  
                    - \del_{xx} \rho^\eps \bigg] \,,
\\
  \del_t \mu^\eps 
  + \del_x \bigg( \pm \frac{3}{2} \rho^\eps \mu^\eps
                 - \frac{3}{4} 
                   \frac{\mu^{\eps\,3}}{\rho^{\eps\,2}} \bigg) 
  & = \frac{\eps^2}{4}
      \del_x \bigg[ \frac{3}{8} {\Cal R}^\eps
                    - \del_{xx} \mu^\eps \bigg] \,, 
\endaligned
  \tag 1.9
$$
where 
$$
  {\Cal R}^\eps = \frac{\del_x \rho^\eps}{\rho^\eps} \del_x \mu
                  - \frac{\del_{xx} \rho^\eps}{\rho^\eps} \mu
                  + \frac{(\del_x \rho^\eps)^2}{2 \rho^\eps} \mu \,.
$$

   The solutions of either the NLS equations or mKdV equations can
then be determined by $\rho^\eps$ and $\mu^\eps$ up to a constant
phase.  The other conserved densities can be also represented in terms of
$\rho^\eps$ and $\mu^\eps$; for example, the energy density
$\epsilon^\eps$ for the focusing $(-)$ and defocusing $(+)$ NLS
hierarchy, defined by
$$
   \epsilon^\eps = \frac{\eps^2}{2} |\del_x u^\eps|^2 
    \mp \frac{1}{2} |u^\eps|^4 \,,
  \tag 1.10
$$
can be determined if $\rho^\eps$ and $\mu^\eps$ are known because
$$
   \vareps^\eps = \frac{1}{2} \frac{|\mu^\eps|^2}{\rho^\eps} 
                  + \frac{\eps^2}{8}
                    \frac{|\del_x \rho^\eps|^2}{\rho^\eps} 
                    \mp \frac{1}{2} \rho^{\eps\,2} \,.
$$
There are recursion formulas (see equation (7.5)) from which all of the
NLS conserved densities can be explicitly generated by iteration. 

\bigskip
\noindent
{\bf 1.3. The Zero-Dispersion Limit.}
We want to understand the behavior of solutions of the ZS hierarchy
when $\eps \to 0$.  Formally, a natural guess is that the limits of
$\rho^\eps$ and $\mu^\eps$, denoted by $\rho$ and $\mu$, would satisfy
the following equations associated to the focusing $(-)$ and
defocusing $(+)$ NLS (1.4a)
$$
\aligned
  \del_t \rho + \del_x \, \mu & = 0 \,, 
\\
  \del_t \mu 
  + \del_x \bigg( \frac{\mu^2}{\rho} \mp \frac{1}{2} \rho^2 \bigg) 
  & = 0 \,;
\endaligned
  \tag 1.11
$$
or the following equations for focusing $(+)$ and defocusing $(-)$
mKdV, (1.6), 
$$
\aligned
  \del_t \rho 
  + \del_x \bigg( \pm \frac{3}{4} \rho^2 
                  - \frac{3}{4} \dfrac{\mu^2}{\rho} \bigg) = 0 \,,
\\
  \del_t \mu 
   + \del_x \bigg( \pm \frac{3}{2} \rho \mu 
                   - \frac{3}{4} \frac{\mu^3}{\rho^2} \bigg) = 0 \,,
\endaligned
  \tag 1.12
$$
with initial values which are the limits of the initial $\rho^\eps$
and $\mu^\eps$.  We call them zero-dispersion limits because the
formally small terms are dispersive.  We call them semiclassical limits
because these limits in general exist only in a weak sense.

Consider the equations for $\rho$ and $\mu$ given by (1.12) with the
focusing sign.  They are the formal limits of the first two
conservation laws specialized to the {\it focusing} mKdV flow (1.9). In the
case that $\mu \equiv 0$, the equation for $\rho$ is nothing but
the Hopf equation (1.2a) which governs the zero-dispersion limit of
the KdV equation (1.1a) near the initial time, except for a trivial
scaling: the time variable $t$ in (1.12) is replaced by $4t$ in (1.2a).  

With a brief calculation, one can find
that the systems (1.11) and (1.12) are not always well-posed for initial
value problems.  They are both hyperbolic for the defocusing case,
which is well-posed; but both elliptic for the focusing case, which is
ill-posed for general initial data.  This has been a principal
obstacle to the further study of the dispersive limit of the focusing
hierarchy.  However, as we shall describe, there is a subset of
the flows, namely the odd flows such as (1.12), which will propagate real 
data ($S\equiv 0$) preserving its reality. (This invariance is decidedly
untrue for the general even flow; for instance, in the focusing NLS flow
itself.) We will demonstrate how the analysis of the dispersive limit,
on this invariant subspace, can be accomplished.

Motivated by the results presented here, Kamvissis, McLaughlin, and
Miller [KMM] very recently undertook a careful analysis of the
dispersive limit of focusing NLS for a very particular initial
condition: $A=\,\hbox{sech}(x)$ and $S\equiv0$.  For this data one has
explicit formulae for the WKB scattering data.  Using newly developed
Riemann-Hilbert problem techniques, they are able to show that the
asymptotic behavior of this particular evolution problem is correctly
described by the Whitham equations, despite their ellipticity.  Given
this development, it now seems probable that this conclusion may hold
more generally for analytic data.

\bigskip
\noindent
{\bf 1.4. Outline of the Paper.}  
In Section 2, we display several numerical experiments to compare the
different situations: well-posed and ill-posed system, complex valued
and real valued system as well as the dynamics among the KdV, focusing
mKdV and the defocusing mKdV initial value problems with the small
parameter $\eps$.  In Section 3, we describe the direct scattering
data of the focusing NLS equation given by (1.4) and its hierarchy.
In Section 4, we describe, via WKB asymptotics, the class of ``single lobe'' 
real initial data; this asymptotic description can be given in terms of 
reflectionless (multi-soliton) potentials.  We analyze
the formulae for $N$-soliton potentials and deduce asymptotic upper and lower 
bounds on these expressions as well convexity properties that enable us to 
establish a KdV-like limit.  In Section 5 we present the rigorous derivation
of this limit and its relation to a fundamental maximization problem. We then
show how the densities and fluxes for the odd flows on real initial data can 
be directly expressed in terms of the maximizer of this problem.
Section 6 then describes the limiting dynamics of these flows in terms of a 
Riemann-Hilbert boundary value problem for the maximizer. Finally, in 
Section 7, we introduce a ``loop algebra'' description of the phase space 
for the NLS system which emphasizes the Lax pair structure of this system.  
In this setting it becomes easy to understand the behavior of the odd flows 
when one restricts to real initial data. One can identify an invariant subspace
for these odd flows, even at finite $\eps$, which further collapses,
formally as $\eps\to0$ onto a space isomorphic to the KdV dispersionless 
densities.  This gives a different approach to the conclusions found in the 
previous sections.  Finally, we make some concluding remarks in Section 8.



\bigskip
\bigskip
\centerline{\bf 2.  Numerical Experiments}
\smallskip

Having introduced the problem, we now present a few numerical
calculations that will serve to illustrate and motivate the 
forthcoming analysis. We will be considering the real initial value
problem for focusing mKdV.

Figure (2.1) presents a sequence of three snapshots of the 
evolution of the focusing mKdV flow (1.6a-b), for the real, 
periodic initial data
$$
	u(x,0)=\cos^2(x)\,, \tag 2.1
$$
with $\eps=0.04$. In these figures one can see quite clearly the 
formation of a region of oscillations, of bounded amplitude, 
within the profile. A fourth panel in this figure shows a space-time 
contour plot of the evolution. For small times one sees that this evolution 
appears to move along straight line characteristics and that the 
oscillations begin to set in where characteristics focus and begin to form 
an envelope. Indeed this evolution very closely matches that of the 
invsicid Burgers equation (1.6a with $\eps = 0$) for the same initial data. 
From the method of characteristics one calculates the break time for this 
dispersionless equation to be $t_B \approx 0.49$, which agrees well with 
where one sees the onset of oscillations. After this time, the solution 
develops a modulated periodic wave train (a so-called ``Dispersive Shock'') 
which propagates away from the initial kink.  Eventually the shock
regions collide creating more complex modulational structures. 

Figure 2.2 shows the evolution of the same initial data as $\eps$ is 
varied between $\eps=0.08$, $\eps=0.04$ and $\eps=0.02$.
Notice the structure of the shocks as $\eps$ is varied.  It appears that 
the width, height, and speed of the modulational envelopes remain 
essentially the same as $\eps \rightarrow 0$, while the oscillations 
within a given envelope are roughly proportional to $1/ \eps$. Because 
of this increasingly oscillatory structure, one does not expect to see 
pointwise convergence, as $\eps \to 0$, of the conserved densities (1.7, 1.10) 
mentioned in the previous section.    

However, Figure (2.3) indicates that local averages of these densities might
have a regular limit: the anti-derivatives of the mass density $\rho^\eps$ 
and of the energy density $\epsilon^\eps$ appear to be converging pointwise
to a limit with smaller $\eps$ producing higher frequency but smaller 
magnitude deviations. Note that $\mu^\eps(x,t)=\eps \text{ Im} 
\, (u^*u_x) =0$ for real initial data in the odd flows.

We conclude this section by mentioning how these simulations were done.
Time was discretized using implicit midpoint or implicit-Runge-Kutta
based on the fourth order Gauss points.  Space was discretized
pseudo-spectrally, with the nonlinear terms computed without
de-aliasing, which was not necessary because we had adequately
resolved the spectrum of the nonlinear terms.  All these calculations
were done in 16-digit precision, and the simulations preserved the
first three conserved quantities to 7-digit accuracy.

%
%
\bigskip
%
%
{%
	\def\x{2.5in}\def\y{2.5in}
	\def\axes#1{%
		\fitclipfalse
		\fitbox[12bp 12bp 228bp 228bp]%
		\fitxyt[2.5in 2.5in 0]%
		\hbox to 0in{\fitfig{figs/#1.eps}}}

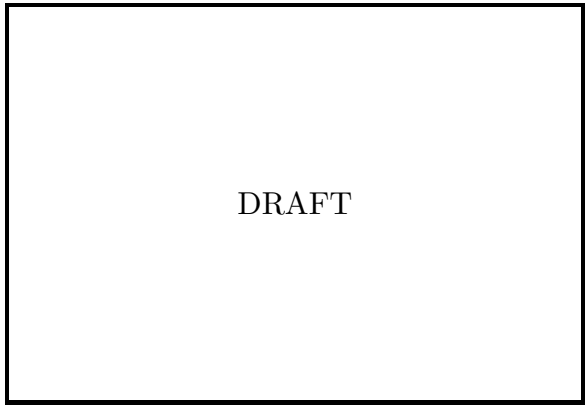
        \def\plaincp#1{%
		\vbox to 2.5in {\hbox to 2.5in {\hfil 
			\fitbox[45bp 45bp 566bp 566bp]%
			\fitxyt[2.5in 2.50in 0]%
			\fitcliptrue
			\fitgz{figs/#1.ps.gz}}\vfil}}
	\def\cp#1#2{%
		\vbox{\hbox{
 		\vbox to \y {\vfil \hbox to \x {%
			\axes{#1}%
			\plaincp{#1}\hfil}}%
		}\vbox to 1.5em{\vfil\hbox to 2.5in{\hfil {#2}\hfil}}}}
	\def\slice#1#2{%
		\fitbox[100bp 100bp 566bp 566bp]%
		\fitxyt[2.5in 0in 0]%
		\fitcliptrue%
		\vbox{\hbox{\fitgz{figs/#1.ps.gz}}\vbox to 1.5em{\vfil\hbox
		to \x{\hfil {#2}\hfil}}}}
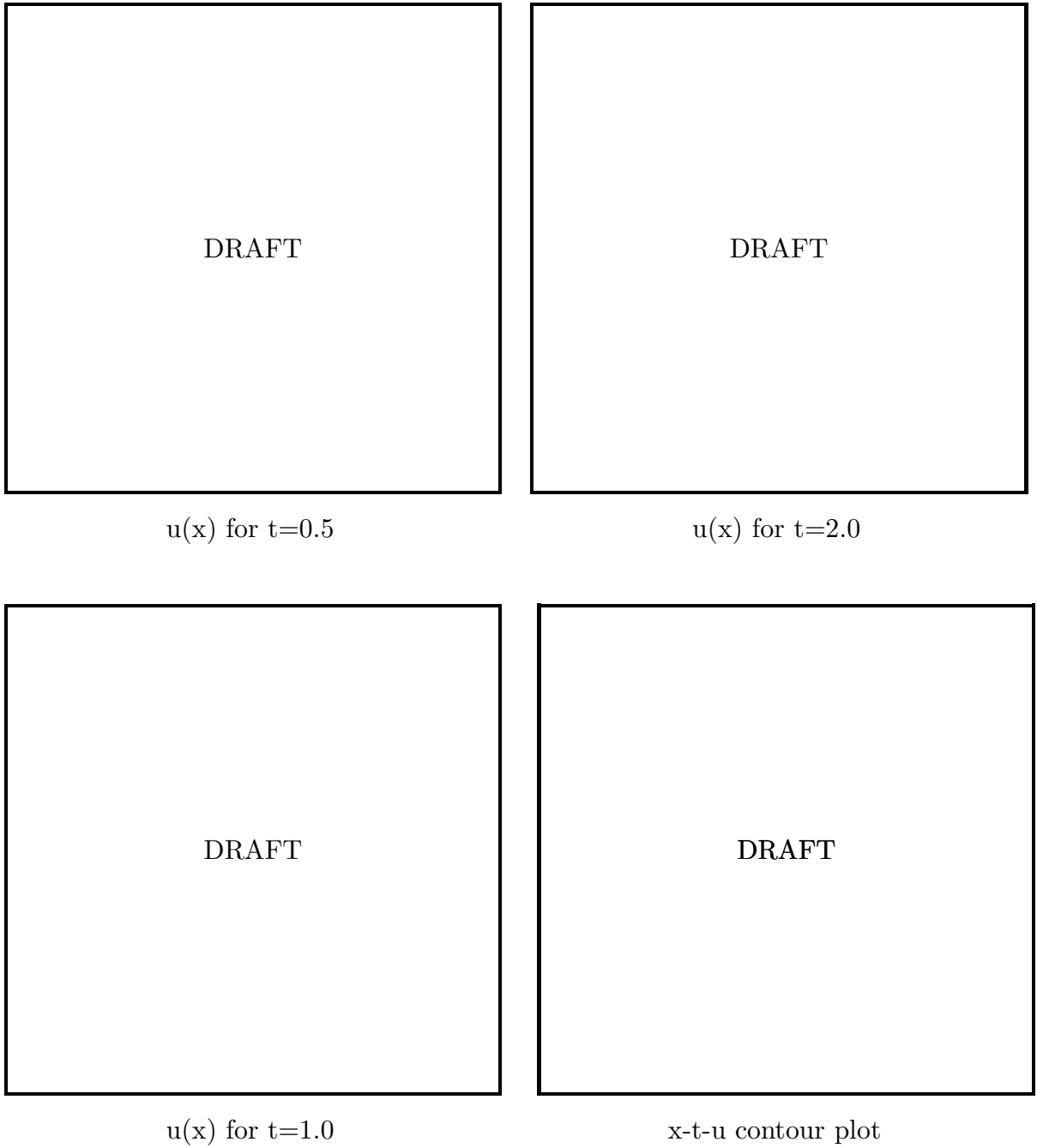
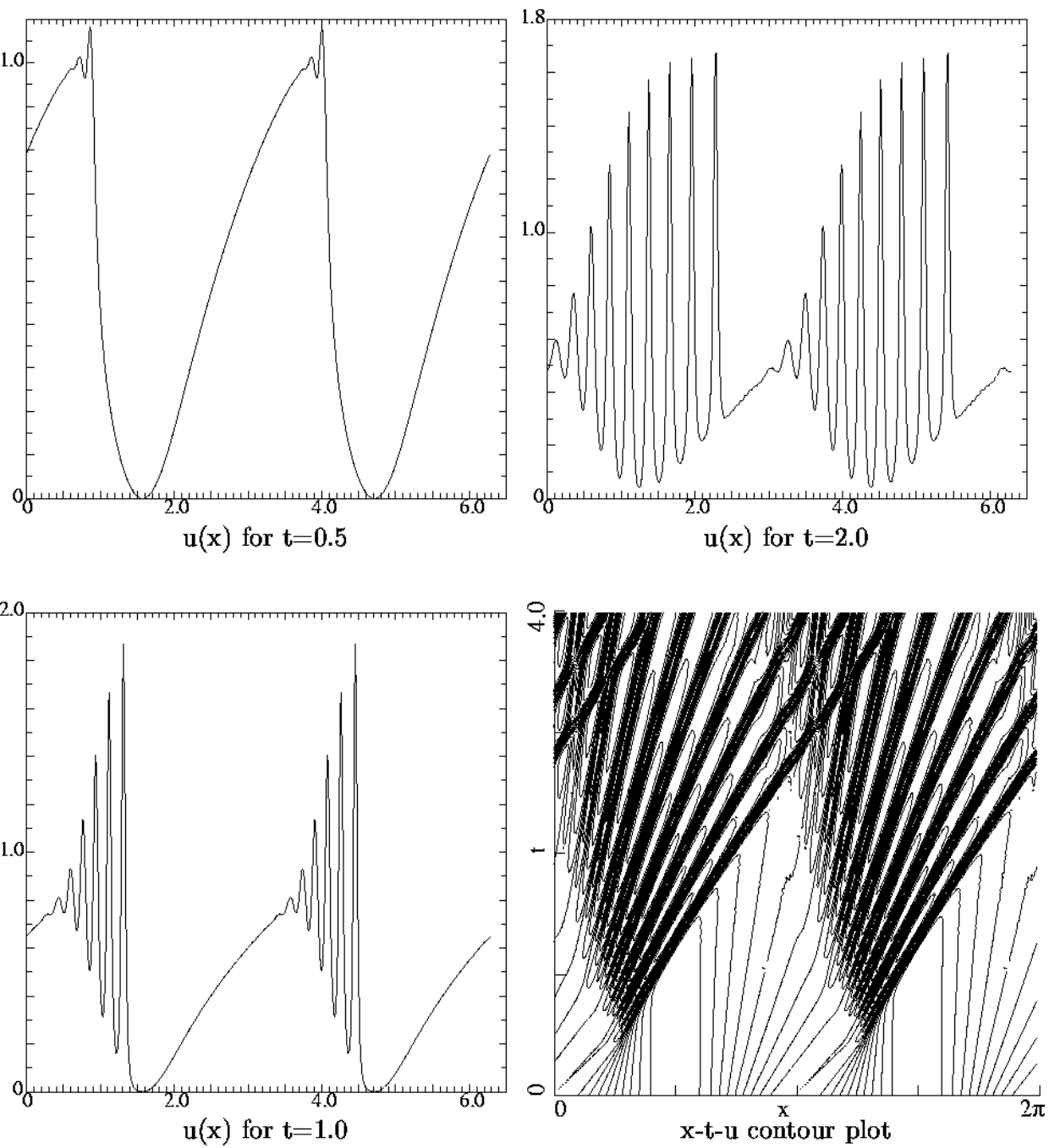
	\def\caption{{\bf Fig 2.1\ } Time-slices, $u(x,t)|_{t=0.5,1.0,2.0}$,
		and contour plot (lower right) $u(x,t)=\text{ const.}$,
		$0<x<2\pi$,$0<t<4$. $u(x,0)=\cos^2 x$.  The horizontal axis
		is $x$, and the vertical axis is either $u$ (slice), or $t$
		(contour plot).}
	\figure{\slice{u_04/u_04s0_5}{u(x) for t=0.5} &
		\slice{u_04/u_04s2_0}{u(x) for t=2.0} \cr
		\vbox to 2em{}\hbox to 8pt{} & \vbox to 2em{}\hbox to 8pt{} \cr
		\slice{u_04/u_04s1_0}{u(x) for t=1.0} &
	\cp{u_04/u_04cp4_0}{x-t-u contour plot}}
}

\bigskip
%
%
\bigskip
%
%
{%
	\def\x{2.5in}
	\def\y{1.75in}
	\def\axes{%
		\fitclipfalse
		\fitbox[12bp 12bp 228bp 228bp]%
		\fitxyt[2.5in 1.75in 0]%
		\hbox to 0in{\fitfig{figs/cp1_5.eps}}}
	\def\plaincp#1{%
		\fitbox[45bp 244bp 566bp 368bp]%
		\fitxyt[2.5in 1.75in 0]%
		\fitcliptrue
		\fitgz{figs/#1.ps.gz}}%
        \def\plainaxes#1{\hbox to 0in{
		\fitbox[0bp 0bp 201bp 202bp]%
		\fitxyt[2.5in 1.75in 0]%
		\fitfig{figs/#1.eps}}}
	\def\cp#1#2{%
		\vbox{\hbox{
 		\vbox to \y {\vfil \hbox to \x {%
			\axes%
			\plaincp{#1}\hfil}}%
		}\vbox to 1.5em{\vfil\hbox to 2.5in{\hfil {#2}\hfil}}}}
	\def\slice#1#2{%
		\fitbox[100bp 100bp 566bp 566bp]%
		\fitxyt[2.5in 1.75in 0]%
		\fitcliptrue%
		\vbox{\hbox{\fitgz{figs/#1.ps.gz}}\vbox to 1.5em{\vfil\hbox
		to \x{\hfil {#2}\hfil}}}}
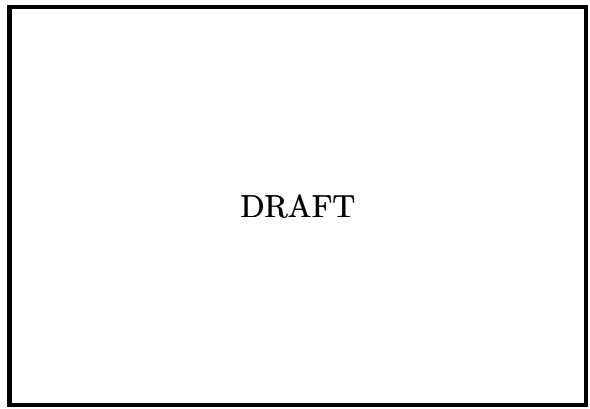
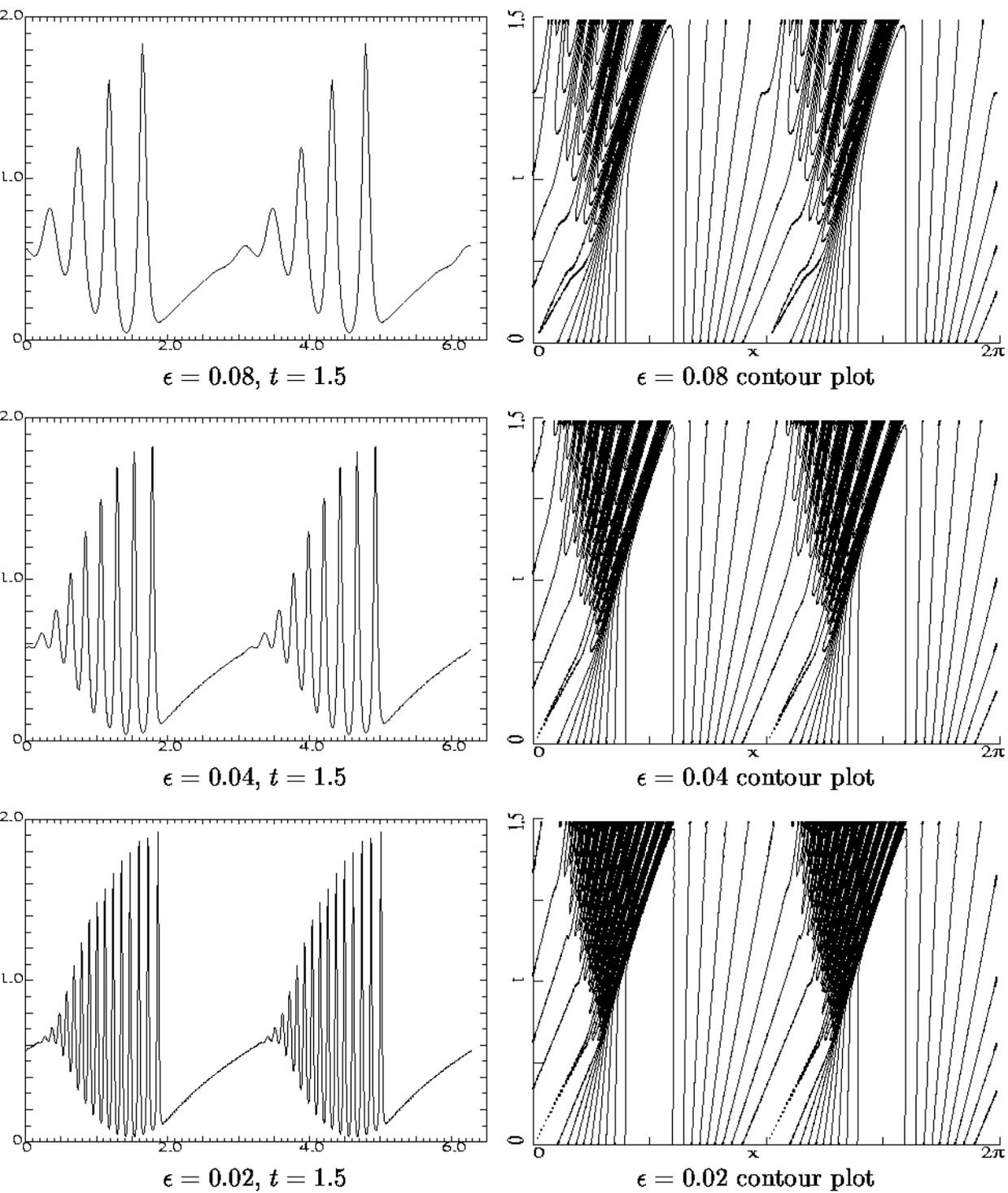
	\def\caption{{\bf Fig 2.2.\ }Contour plot,
	$u_\eps(x,t)=\text{ const.}$, $ 0<t<1.5$ (left). Slice,
	$u_\eps(x,t)|_{t=1.5}$, (right). Here, $\eps=0.08$ (top),$0.04$
	(center),and $0.02$ (bottom), and $u_\eps(x,0)=\cos^2 x$. The
	horizontal axis is $x$, and the vertical axis is either $t$
	(contour plot) or $u$ (slice).  The graphs have the same scales
	for comparison. }
	\def\thisfigure#1{\vbox{
	\centerline{\vbox{$$\matrix #1 \endmatrix$$}}
	\vskip -0.5em
	\noindent{\caption}}}
	\thisfigure{\slice{u_08/u_08s1_5}{$\eps=0.08$, $t=1.5$} &
		\cp{u_08/u_08cp1_5}{$\eps=0.08$ contour plot} \cr
		\hbox to 8pt{} & \hbox to 8pt{} \cr
		\slice{u_04/u_04s1_5}{$\eps=0.04$, $t=1.5$} & 
		\cp{u_04/u_04cp1_5}{$\eps=0.04$ contour plot} \cr
		\hbox to 8pt{} & \hbox to 8pt{} \cr
		\slice{u_02/u_02s1_5}{$\eps=0.02$, $t=1.5$} & 
		\cp{u_02/u_02cp1_5}{$\eps=0.02$ contour plot}}
}

\bigskip
%
%
\bigskip
%
%
{%
	\def\x{2.25in}\def\y{2.25in}
	\def\spec#1{%
		\fitbox[100pt 100pt 566pt 566pt]%
		\fitxyt[2.25in 2.25in 0]%
		\fitcliptrue
		\fitspecial{"`gunzip -c figs/#1.ps.gz"}}%
	\def\overlay#1#2{%
		\iffitdraft
			\fitbox[0in 0in 1in 1in]%
			\fitxyt[2.25in 2.25in 0]%
			\fitfig{#1}%
		\else
			\vbox to \y {\vfil\hbox to \x{%
			\spec{#1}\spec{#2}\hfil}}%
		\fi}%

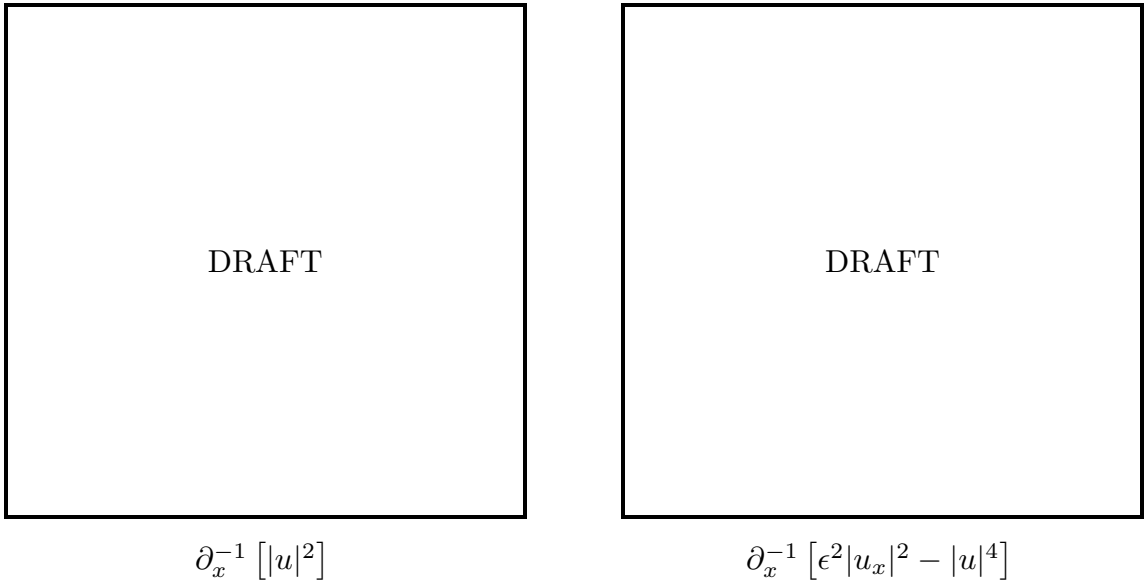
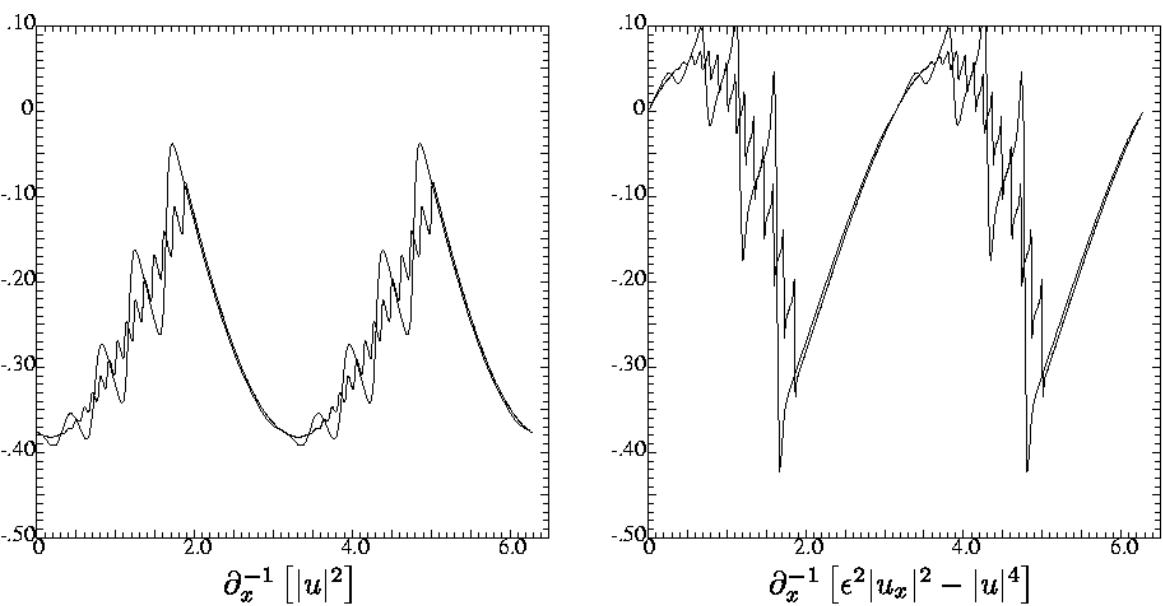
	\def\caption{{\bf Fig 2.3\ } Primitives of the first (left)
	and third (right) conserved densities.  The cases for for
	$\eps=0.08$ and $\eps=0.02$, are overlaid, with the wider
	oscillations corresponding to large $\eps$. Here, the
	pseudo-inverse $\partial_x^{-1}$ is a periodic normalization
	of $\int^x$: each discrete Fourier mode of the density is
	multiplied by $\frac{1}{ik}$, except for mode 0, which is
	left unchanged.  As $\eps$ decreases, both the wavelength and
	the magnitude of the deviating oscillations decrease.  This
	suggests that the limit for the primitives will be classical.}

	\figure{%
		\vbox{\vbox{\hbox{\overlay{u_02/u_02fi1_5}{u_08/u_08fi1_5}}}
		\vbox to 1.5em{\vfil\hbox to \x{\hfil {$\partial_x^{-1} \left[|u|^2\right]$}\hfil}}}&
		\hbox to 1em{} &
		\vbox{\vbox{\hbox{\overlay{u_02/u_02fiii1_5}{u_08/u_08fiii1_5}}}
		\vbox to 1.5em{\vfil\hbox to \x{\hfil {$\partial_x^{-1} \left[\eps^2 |u_x|^2-|u|^4\right]$}\hfil}}}}
}

\bigskip
%
%



\bigskip
\bigskip
\centerline{\bf 3. The Focusing Zakharov-Shabat Hierarchy}
\smallskip

\bigskip
\noindent
{\bf 3.1. The Scattering Transform.} 
Zakharov and Shabat [ZS1] solved the Cauchy problem for the focusing
NLS equation (1.4) with initial data $A(x)$ and $\del_x S(x)$ that
decay sufficiently rapidly as $|x|\to\infty$.  They used the inverse
scattering transform associated 
with the nonselfadjoint Dirac operator
$$
  \LL 
  = \pmatrix -i \eps \, \del_x  &      i \bar{u}     \\
                     i u        &  i \eps \, \del_x  \endpmatrix \,.
  \tag 3.1a
$$
Here we collect the relevant facts regarding their theory.  Throughout
this section $\eps>0$ will be considered to be fixed, and hence, no 
explicit $\eps$ dependence will be indicated.  

   The solution strategy centers on the spectral problem
$$
  \LL f = \zeta f \quad
  \hbox{where} \quad
  f = \pmatrix  f^{(1)} \\ f^{(2)} \endpmatrix \,.
  \tag 3.1b
$$
Given $u(x,0)$, the asymptotics of the eigenfunctions $f(\zeta,x,0)$
as $|x|\to\infty$, referred to as the scattering data, can be
calculated in principle.  The evolution of the scattering data is
governed by simple formulas and $u(x,t)$ is then obtained from the
knowledge of the large $|x|$ asymptotics of $f(\zeta,x,t)$ using
inverse scattering theory.

   More specifically, if $u(x,0)$ has the form (1.4b) where $A(x)$ and
$\del_x S(x)$ decay sufficiently rapidly as $x\to\pm\infty$ (say,
faster than any power of $x$), then the $L^2$ spectrum of $\LL$
consists of the whole real axis, comprising the continuous spectrum,
along with a finite set (possibly empty) of simple eigenvalues in the
complex plane.  It is easily seen that if $f$ and $\zeta$ satisfy (3.1)
then so do $\tilde{f}$ and $\bar{\zeta}$ where 
$$
  \tilde{f} = \pmatrix  \bar{f}^{(2)} \\ 
                       -\bar{f}^{(1)} \endpmatrix \,.
  \tag 3.2
$$
We will therefore only consider $\zeta_1,\cdots,\zeta_N$, the $N$
eigenvalues of $\LL$ located in the upper half of the complex
plane.


   To describe the asymptotic behavior of the $L^2$ scattering data we first
introduce the Jost fundamental solution $F(\zeta,x)$.  For each
$\zeta\in\C$, this matrix solution is uniquely defined by the
following two conditions:
$$
\aligned
  i) & \qquad F(\zeta, x) \exp(-i \sigma_3 \zeta x/\eps) \to \I \,, 
       \qquad x \to +\infty
\\
  ii) & \qquad \sup_{x \in \R} \| F(\zeta, x) \| < \infty \,,
\endaligned
$$
where $\sigma_3 = \pmatrix 1 & 0 \\ 0 & -1 \endpmatrix$.  This Jost
solution is meromorphic in the open upper half-plane, as well as in the 
open lower half-plane, and has limiting values,
$$
  F_\pm(\lambda,x)  = \lim_{\eps\to 0}F(\lambda\pm i\eps, x) \,,
$$
from above and below respectively, on the real axis $\lambda \in \R$, which
are continuous in $\lambda$. These two limiting values are related by a 
``jump matrix'' as follows:
$$
F_+(\lambda) = F_-(\lambda)\pmatrix 1+|R(\lambda)|^2 & \BAR{R(\lambda)} \\
R(\lambda) & 1 \endpmatrix,
$$
where the function $R(\lambda)$ appearing in the jump matrix is called
the {\it reflection coefficient}. It is complex-valued and depends only on
$\lambda$.

The first and second columns of $F(\zeta, x)$, 
respectively denoted by $f_1$ and $f_2$, have the following asymptotics in 
$\zeta$:
$$
\aligned
  f_1(\zeta,x) & = \exp\!\bigg( \dfrac{i \zeta x}{\eps} \bigg)
                   \left[ \pmatrix 1 \\ 0 \endpmatrix 
                          + \OO (\zeta^{-1}) \right] \qquad 
  \text{for $\zeta\to\infty$} \,,
\\
  f_2(\zeta,x) & = \exp\!\bigg( \dfrac{-i \zeta x}{\eps} \bigg)
                   \left[ \pmatrix 0 \\ 1 \endpmatrix 
                          + \OO (\zeta^{-1}) \right] \qquad 
  \text{for $\zeta\to\infty$} \,,
\endaligned
$$
The eigenvalues $\zeta_j$ in the upper half-plane are determined by
the condition that these two solutions should be linearly dependent.
(These are also precisely the locations where the Jost function in the
upper half plane has a pole. Away from these points the Jost function is
analytic.)  

At an eigenvalue, the proportionality between these two solutions
$$
  f_1(\zeta_j,x) 
  = \exp\!\bigg( - \dfrac{\chi_j}{\eps} \bigg) f_2(\zeta_j,x) 
$$
is specified in terms of $\chi_j$, which is referred to as the norming
exponent.

   The inverse theory prescribes that the fundamental scattering data
consists of the reflection coefficient $R(\lambda)$, the $N$ eigenvalues
$\zeta_j$ and the $N$ norming exponents $\chi_j$.  

   As $u(x,t)$ evolves according to the focusing NLS equation (1.4),
the eigenvalues $\{\zeta_j\}$ remain unchanged while the time
dependence of the other scattering data is
$$ 
  \chi_j(t) = \chi_j + i \zeta^2_j t \,, \qquad 
  R(\lambda,t) = R(\lambda) \exp\!\bigg( \frac{2 i \lambda^2 t}{\eps} \bigg) \,.  
  \tag 3.3
$$ 
Given $R(\lambda)$, $\zeta_j$, and $\chi_j$ computed from the
initial data $u(x,0)$, the solution $u(x,t)$, as well as all of the conserved
densities defined in the next subsection, of the NLS equation (1.4),
are then determined by inverse scattering from $R(\lambda,t)$,
$\zeta_j$, and $\chi_j(t)$ given by (3.3) [ZS1, KMM].

\bigskip
\noindent
{\bf 3.2. The Hierarchy.} 
The complete integrability of this cubic \Schrodinger equation implies
the existence of an infinite family of independent, conserved
quantities [ZS1],
$$
   H_m = \int_{-\infty}^\infty \rho_m \, dx \qquad 
   \hbox{for $m=0,1,\cdots$} \,,
$$
which are in involution with respect to the Poisson bracket,
$$
  0 = \{H_m, H_n\} 
    = \frac{1}{i \eps} 
       \int_{-\infty}^\infty
           \bigg( \frac{\delta H_m}{\delta u} 
                  \frac{\delta H_n}{\delta \bar{u}}
                  - \frac{\delta H_m}{\delta \bar{u}} 
                    \frac{\delta H_n}{\delta u} \bigg) \, dx \,.
$$
The first three densities $\rho_m$ are essentially those given by
(1.7) and (1.10):
$$
\aligned
  \rho_0 & = - \rho = - |u|^2 \,, 
\\
  \rho_1 & = \mu = -i \frac{\eps}{2} 
                   \left( \bar{u} \del_x u 
                          - u \del_x \bar{u} \right) \,, 
\\
  \rho_2 & = \frac{3}{2} \vareps^\eps 
           = \frac{3}{4} 
             \left( \eps^2 |\del_x u|^2 - |u|^4 \right) \,. 
\endaligned
  \tag 3.4
$$
A recursion formula that generates all the conserved densities is
given by (7.5).

   All of the $H_m$ are Hamiltonians that generate flows which
commute with the focusing cubic NLS flow (1.4a), the so-called
focusing Zakharov-Shabat (ZS) hierarchy.  Letting $t_m$ denote the
time variable associated with the $m^{th}$ flow, its evolution is then
given by
$$
  i \eps \, \del_{t_m} u = \frac{\delta H_m}{\delta \bar{u}} \qquad
  \text{for $m=0,1,\cdots$} \,. 
$$
The $t_0$ flow is just a phase rotation, the $t_1$ flow is just a
spatial translation, the $t_2$ flow is given by the focusing NLS
equation (1.4a) with $t=t_2$, while the $t_3$ flow is that of the
complex mKdV equation
$$
  \del_{t_3} u + \frac{3}{2} |u|^2 \del_x u 
  + \eps^2 \frac14 \del_{xxx} u = 0 \,.
$$
Every $H_n$ is conserved by each flow; their densities satisfy the
local conservation laws 
$$
\aligned
  \del_{t_0} \rho_{n-1} & = 0 \,,
\\
  \del_{t_m} \rho_{n-1} + \del_x \mu_{m,n} & = 0 \,, \qquad
\endaligned 
  \text{for $m,n=1,2,\cdots$} \,. 
  \tag 3.5
$$
Here $\mu_{m,n}$ is the flux for the $(n-1)^{st}$ conserved
quantity under the $m^{th}$ flow. 

   Because all these flows commute, they may be solved simultaneously
for $u(x,\tt)$, having the form of the initial condition (1.4b),
where $\tt=(t_0,t_1,\cdots)$ such that all but finitely many $t_m$ are
zero.  Associated with each $\tt$ is a polynomial $p(\,\cdot\,,\tt)$
defined by
$$ 
  p(\zeta,\tt) = \sum_{m=0}^\infty t_m \zeta^m \,.
  \tag 3.6
$$ 
The simultaneous evolution of the scattering data is then given by 
$$
  \chi_j(\tt) = \chi_j + i p(\zeta_j,\tt) \,, \qquad 
  R(\lambda,\tt) = R(\lambda) 
               \exp\!\bigg( \frac{ 2 i p(\lambda, \tt)}{\eps} \bigg) \,,
  \tag 3.7 
$$ 
and $u(x,\tt)$ is determined by inverse scattering.  

   In Section 7 we show that the odd flows in this hierarchy preserve
real symmetry.  More specifically, we show that $u(x,\tt)$ is real 
whenever $u(x,0)$ is real and $t_m=0$ for every even $m$.  We show
moreover that when $u$ is real that $\rho_n=0$ for every odd $n$.    

   The scope of the zero-dispersion limit for the NLS equation can
then be enlarged to consider the solution $u^\eps(x,\tt)$ of the whole
hierarchy that satisfies the initial condition
$$
  u^\eps(x,0) 
  = A(x) \exp\!\bigg( \frac{i S(x)}{\eps} \bigg) \,, 
$$
where $A(x)$ is positive, $S(x)$ is real, both are independent of
$\eps$, and $A(x)$ and $\del_x S(x)$ are Schwartz class.  The goal
is then to determine the limiting behavior of all the conserved
densities $\rho_n^\eps$ and fluxes $\mu_{m,n}^\eps$ associated with
the entire focusing ZS hierarchy of flows as $\eps$ tends to zero.
In this paper we achieve this goal only for the case when $u$ is real
and we restrict to the odd flows of the hierarchy.

\bigskip
\bigskip
\centerline{\bf 4. Analysis of the Scattering Transform} 
\smallskip

\bigskip
\noindent
{\bf 4.1. Asymptotic Analysis of the Initial Scattering Data.}
For initial data 
$$
  u^\eps(x,0) 
  = A(x) \exp\!\bigg( \frac{i}{\eps} S(x) \bigg) \,,
$$
of the class described at the end of the previous section, one can
asymptotically determine the initial scattering data by a WKB
analysis.

The $L^2$ eigenvectors as well as the reflection coefficient, 
$R(\lambda)$, for the Zakharov-Shabat eigenvalue problem (3.1) were 
defined in Section 3. The WKB ansatz is that the 
eigenvectors have a leading order asymptotic (in $\epsilon$) behavior 
of the form
$$
  f^{\eps\,(1)} 
  ={r}^{(1)} \exp\!\left( \frac{i (\theta - \frac12 S)}
                              {\eps} \right) + \cdots \,, \qquad
  f^{\eps\,(2)} 
  = {r}^{(2)} \exp\!\left( \frac{i (\theta + \frac12 S)}
                              {\eps} \right) + \cdots \,, 
$$
where ${r}^{(1)}$, ${r}^{(2)}$, and $\theta$ do not depend on $\eps$.
(There is a similar ansatz for the Jost functions.)
When this ansatz is substituted into $\LL f^\eps=\zeta f^\eps$, the leading
order gives
$$
  \pmatrix \zeta + (\del_x \theta - \frac12 \del_x S) &  
           i A                                        \\
           i A                                        &  
           \zeta - (\del_x \theta + \frac12 \del_x S) \endpmatrix
  \pmatrix {r}^{(1)} \\ {r}^{(2)} \endpmatrix = 0 \,.
$$
This implies that 
$$
\aligned
  0 & = \det\!\pmatrix \zeta + (\del_x \theta - \frac12 \del_x S) &  
                       i A                                        \\
                       i A                                        &  
                       \zeta - (\del_x \theta + \frac12 \del_x S) 
                       \endpmatrix
\\
    & = (\zeta - \tfrac12 \del_x S)^2 - (\del_x \theta)^2 + A^2 \,.
\endaligned
$$

   Suppose now that the initial value of $u_\eps$ is real.  That means
$u_\eps(x,0)=A(x)$ and 
$$
  (\del_x \theta)^2 = (A^2 + \zeta^2) \,. 
$$
If the WKB asymptotics are valid then, in order for $f^{\eps}$ to be an
eigenvector, $\del_x\theta(x)$ should be pure imaginary for sufficiently large
values of $|x|$. It follows from the preceding equation that $\zeta$ should
then be pure imaginary. Thus, we would expect the point spectrum to converge, 
as $\eps \to 0$, onto the imaginary axis. 

In fact there is a natural class of initial data, the so-called ``single lobe''
potentials for which the eigenvalues are always pure imaginary independent of
the value of $\eps$. By a single lobe potential we shall mean a real-valued
potential for which $A(x)$ is nonegative, smooth and integrable on $\R$ and
moreover $A(x)$ should be non-decreasing on the left of $x=0$ and 
non-increasing on the right. It has recently been demonstrated [KS] that for 
such data the eigenvalues are {\it always} pure imaginary. In fact these 
authors show that this result holds under a somewhat weakened definition 
of single lobe, but the above will suffice for our purposes.   
For such data, $A(x)$ will have two turning points, $\Xpm(\eta)$ (Fig. 4.1).  
We will assume that $A(x)$ has maximum height 1 . In this setting the WKB 
ansatz implies that the asymptotic scattering data should satisfy:
 
\item{(1)} The reflection coefficient $R(\lambda) \equiv 0$.

\item{(2)} The eigenvalues are independent of time.  They are purely 
imaginary with $\zeta_j = i \eta_j$. The density of states, for the 
positive imaginary eigenvalues, is 
$$
\phi(\eta) = \frac{1}{\pi} \frac{d}{d\eta}
\REAL \int \sqrt{A^2(y) - \eta^2} \, dy \, = \frac{1}{\pi} \frac{d}{d\eta}
\int_{x_-(\eta)}^{x_+(\eta)} \sqrt{A^2(y) - \eta^2} \, dy \,;
$$
the eigenvalues are then defined by 
$$
  (j+\tfrac{1}{2}) \eps = -\int_{\eta_j}^{\eta_{\max}} \phi(\eta) d\eta
$$
       for $j=1,2,\cdots,N^\eps$ and $\eta_1<\eta_2<\cdots<\eta_{N^\eps}$ 
       where

\item{(3)} $N^\eps$ is the total number of eigenvalues in the upper half
plane and is determined by
$$
  N^\eps = \bigg[ \frac{1}{2\pi \eps} \REAL \int A(y) \, dy \bigg] \,.
$$

\item{(4)} The initial norming exponents can be approximated by
$$
  \chi_j(0) = \chi_j 
 = \eta_j \, \Xp(\eta_j) 
    + \int_{\Xp(\eta_j)}^\infty
          \left( \eta_j - \sqrt{\eta_j^2 - A^2(x)} \right) \, dx \,.
$$


\bigskip
{%
	\def\fig#1{%
		\fitbox[0bp 0bp 608bp 239bp]
		\fitxyt[5in 0in 0]%
		\fitgz{figs/#1.eps.gz}}%
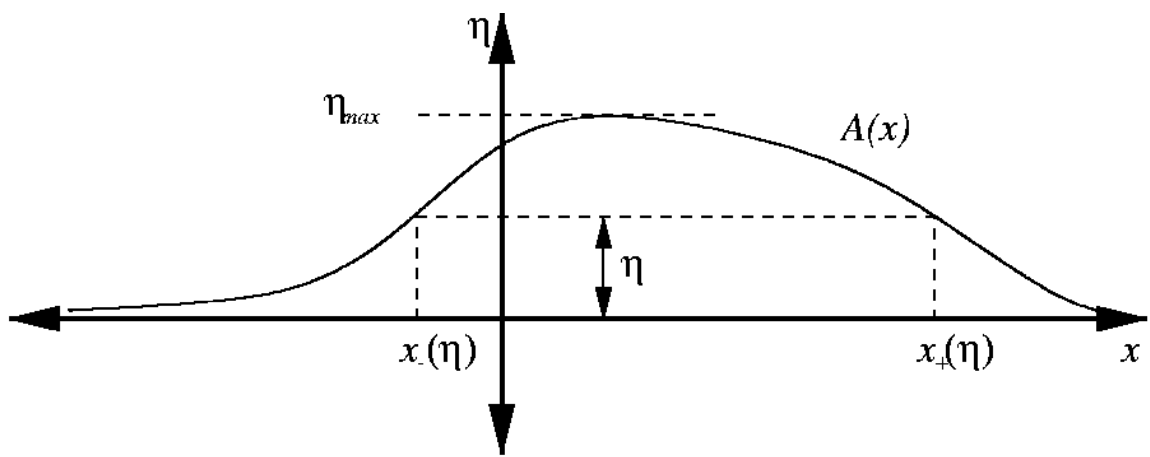
	\def\caption{{\bf Fig 4.1\ }The initial data $A(x)$.  Note its
          critical value $\eta_{max}$ and the indicated defining
          relations for $\Xpm=\Xpm(\eta)$.}
	\figure{\fig{data}}
}%

\bigskip

Thus, if one just considers the odd flows, i.e. $\tt = (-t_1, t_3, \dots 
(-1)^k t_{2k-1}, \dots) \,$ then all the scattering data are real. The 
formula for $\chi_j$ given by (3.7) can be rewritten as
$$
  \chi_j(\tt) 
  = \chi_j + \sum_{k=1}^\infty  \eta_j^{2k-1}t_{2k-1} \,.
  \tag 4.2
$$
We observe that this asymptotic data is the same as that for the
\Schrodinger operator (1.3) with $A^2(x)$ playing the role of
$u_{in}(x)$ [LLV].  Henceforth, we consider the Cauchy problem for
focusing NLS (1.4a) with initial data corresponding to the above
asymptotic scattering data.

\bigskip
\noindent
{\bf 4.2. Reflectionless Approximation.}
Motivated by the previous section, we choose to replace the exact 
real-valued initial data $A$ by the real-valued reflectionless 
potential $A^\eps$ corresponding to the WKB scattering data given
above.  This allows us to use the reconstruction formulas for
reflectionless potentials [FNR] that give the associated conserved
densities and fluxes by
$$ 
\aligned
   \rho_{n-1}^\eps(x,\tt) 
   & =  \eps^2 \del_{x t_n} \log(\tau^\eps(x,\tt)) \,, 
\\ 
   \mu_{m,n}^\eps(x,\tt) 
   & =  - \eps^2 \del_{t_m t_n} \log(\tau^\eps(x,\tt)) \,,
\endaligned
  \tag 4.3
$$
where the so-called tau-function $\tau^\eps(x,\tt)$ is given by the
$N^\eps\times N^\eps$ determinant
$$
  \tau^\eps(x,\tt) 
  = \det\big( I + \eps^2 G^\eps(x,\tt) \BAR{G^\eps(x,\tt)} \big) \,.
  \tag 4.4
$$
Here $I$ is the identity matrix while the $N^\eps\times N^\eps$ matrix
$G^\eps$ has the form
$$
  G^\eps(x,\tt)
  = \Bigg( \frac{i}{\zeta_j - \BAR{\zeta_k}} \,
           \exp\!\bigg( \frac{a(i\zeta_j,x,\tt) 
                              + \BAR{a(i\zeta_k,x,\tt)}}{\eps} \bigg)
           \Bigg)_{j,k=1}^{N^\eps} \,,
  \tag 4.5
$$
where 
$$
  a(\eta,x,\tt) \equiv - \eta x + \chi_j(\tt) \,.
  \tag 4.6
$$

All the eigenvalues $\{\zeta_{j}\}_{1}^{N}$ lie on the
positive imaginary axis and are bounded as follows:
$$
  \zeta_j = i \eta^\eps_j \,, \qquad 
  0 < \eta^\eps_1 < \eta^\eps_2 < \cdots < \eta^\eps_N < 1 \,,
$$
then $E$ is real for all the odd flows ($t= t_1, t_3, \cdots$)
and
$$
\frac{i}{\zeta_j - \BAR{\zeta_k}}  = \frac{1}{\eta^\eps_j + \eta^\eps_k}.
$$
From this, we see that $G^\eps$ is real, positive
definite and symmetric.

In this setting we define
$$
B(x,\tt) = G^\eps(x,\tt) = \BAR{G^\eps(x,\tt)}.
$$

\bigskip
\noindent
{\bf 4.3. Convexity of the Tau-Function.}
We will make use of the following property of the tau-function.

{\narrower
\smallskip
\noindent
{\bf Lemma 4.1.} For each $\eps>0$ the function  
$$ 
  (x,\tt) \mapsto \tau(x,\tt) \quad \hbox{is convex} \,. 
$$

\smallskip}
\noindent 
{\bf Proof.}  Set 
$W = \eps^2 \log \det (I+\eps^2 B^2)$.  It is sufficient to show that
$W$ is convex in $x$ for general real parameters $\eta_j$ and
$\chi_j$.  This is true because convexity is a property along line
segments, and examining $W$ along a general line, parameterized by
$\tilde{x}$, of the form $P(\tilde{x})
=(x(\tilde{x}),\tt(\tilde{x}))=P_0(1-\tilde{x})+P_1\tilde{x}$ 
results in the function 
$$
  W(\tilde{x}) = \eps^2 \log \det (I+\eps^2 B(\tilde{x})^2) \,.
$$
The convexity of $W(x,\tt)$ follows from the convexity of $W(\tilde{x})$.
For the remainder of the argument, we drop the tildes and denote 
differentiation in $\tilde{x}$ with a prime ($'$).

Using the definition of $B$, we find $B'=-\frac{1}{\eps}(\Lambda B
+ B \Lambda)$, where $\Lambda$ is the diagonal matrix of the $\eta_j$.
From this and the general facts that $\det(a)'=\det(a)
(\tr(a^{-1}a')$ and $\tr(ab)=\tr(ba)$, we find by direct calculation
that
$$
  W' = - 4 \eps \, 
         \tr\!\left( \left( I + \eps^2 B^2 \right)^{-1} 
                     \eps^2 B^2 \Lambda \right)\,.
  \tag 4.7
$$

It is more convenient to write $W'$ as 
$$
  W' = - 4 \eps \, 
         \tr\!\left( \left( I + \left( \eps^2 B^2 \right)^{-1} \right)^{-1} 
                     \Lambda \right) 
$$
before differentiating to obtain 
$$
  W'' = 4 \, \tr\ \left[ ( A ) (\Lambda B^2 A \Lambda ) \right] 
        + 8 \, \tr\ \left[ ( A B ) (\Lambda B A \Lambda ) \right]
        + 4 \, \tr\ \left[ ( A B^2 ) (\Lambda A \Lambda ) \right] \,,
  \tag 4.8
$$
where
$$
  A = \left( I + (\eps B)^{-2} \right)^{-1} (\eps B)^{-2} \,.
$$
Both $A$ and $B$ are symmetric positive definite (SPD), and $A$
commutes with $B$, so products of $A$ with powers of $B$ are also SPD.
Thus the terms of $W''$ in (4.8) are traces of the products of two SPD
matrices, each grouped in parentheses in the equation.

   Traces of products of SPD matrices are positive, and so the sum is
positive. Thus $W'' \geq 0$, and so $W(\tilde{x})$ is convex, and, so,
$W(x,\tt)$ is convex, which proves Lemma~4.1.

\bigskip

   We remark that (4.7) for the full system yields
$$
  \eps^2 \del_x \log\tau^\eps(+\infty,\tt) = 0 \,, \qquad 
  \eps^2 \del_x \log\tau^\eps(-\infty,\tt) = -4 \eps\ \tr(\Lambda) \,.
$$
One can similarly find $\eps^2\del_{t_m}\log\tau^\eps(\pm\infty,\tt)$
for any positive, odd number $m$.

\bigskip
\bigskip
\centerline{\bf 5. Establishing the Limit}
\smallskip

\bigskip
\noindent

{\bf 5.1. The Limit of the Tau-Functions.}  
In this subsection we establish the limit, as $\eps$ tends to zero, of
$\eps^2\log\!\big(\tau^\eps(x,\tt)\big)$ in the topology of uniform
convergence over compact subsets of $\R\times\TT_o$, characterizing
the limit in terms the solution of a maximization problem. Here, 
$\TT_o$ denotes the subspace of vectors of the form $\tt = (- t_1, t_3, \dots
(-1)^k t_{2k-1}, \dots)$ which are real and have finite support.
\bigskip

   We adopt the following notation.  Given any $N^\eps\times N^\eps$
matrix $A$ indexed by the set $J^\eps=\{1,\ldots,N^\eps\}$, and given
any two nonempty subsets $S\subset J^\eps$ and $T\subset J^\eps$, let
$A_{ST}$ denote the $|S|\times|T|$ submatrix of $A$ obtained by
retaining only those entries whose row index belongs to $S$ and column
index belongs to $T$.  We abbreviate $A_{SS}$ as $A_S$.  Finally, for
every $S\subset J^\eps$, we let $S'$ denote the compliment of $S$ in
$J^\eps$.

   The key new result is the following.

{\narrower
\smallskip
\noindent
{\bf Theorem 5.1.} Let
$$
  q^\eps(x,\tt) 
  = \max\!\Big\{ \eps^2 
                 \log\big( \det\big( G^\eps_S(x,\tt) \big) \big) \,:\,
                 S \subset J^\eps \Big\} \,,
  \tag 5.1
$$
where the maximum ranges over all index sets, with the understanding
that $\det(G^\eps_S)=1$ when $S$ is the empty set.  Then
$$
  \lim_{\eps\to0} 
  \Big| \eps^2 \log\!\big( \tau^\eps(x,\tt) \big)
        - q^\eps(x,\tt) \Big| = 0 
  \tag 5.2
$$
uniformly over $\R\times\TT_o$.

\smallskip}
\noindent 
{\bf Proof.} Below we will derive the crude bounds
$$
  \eps^{2 N^\eps} \exp\!\bigg( \frac{q^\eps(x,\tt)}{\eps} \bigg)
  \leq \tau^\eps(x,\tt)
  \leq 2^{2 N^\eps} \exp\!\bigg( \frac{q^\eps(x,\tt)}{\eps} \bigg) \,.
  \tag 5.3
$$
Given these, by taking their logarithm, multiplying the result by
$\eps^2$, and subtracting $q^\eps(x,\tt)$, one arrives at
$$
  \eps^2 2 N^\eps \log(\eps) 
  \leq \eps^2 \log\!\big( \tau^\eps(x,\tt) \big) - q^\eps(x,\tt)
  \leq \eps^2 2 N^\eps \log(2) \,.
  \tag 5.4
$$
Because $N^\eps=O(1/\eps)$ by the WKB ansatz, the limit (5.2) follows.
Hence, the theorem follows once the bounds (5.3) are established.

   To derive the upper bound on $\tau^\eps$ in (5.3), first use the
fact that $G^\eps$ is Hermitian positive to obtain
$$
\aligned
   \tau^\eps  
   & = \det\!\big( I + \eps^2 G^{\eps\,2} \big) 
\\
   & \leq \det\!\big( I + 2 \eps \, G^\eps + \eps^2 G^{\eps\,2} \big)
     = \big[ \det\!\big( I + \eps \, G^\eps \big) \big]^2 \,.
\endaligned
  \tag 5.5
$$
The last determinant in (5.5) is just the characteristic polynomial of
$\eps\,G^\eps$ evaluated at $-1$ and can be expanded in terms of the
determinants of its minor matrices $\eps\,G^\eps_S$ as
$$
  \det\!\big( I + \eps \, G^\eps \big) 
  = \sum_{S \subset J^\eps} \eps^{|S|} \det\!\big( G_S^\eps \big) \,.
  \tag 5.6 
$$
Because every principal minor of $G^\eps$ has the same form as
$G^\eps$, each $G^\eps_S$ is Hermitian positive, and hence, has a
positive determinant.  The sum in (5.6) therefore contains
$2^{N^\eps}$ positive terms, each of which is bounded above (for
$\eps\leq1$) by
$$
  \eps^{|S|} \det\!\big( G_S^\eps \big)
  \leq \exp\!\bigg( \frac{q^\eps(x,\tt)}{2 \eps} \bigg) \,,
  \tag 5.7
$$
The upper bound on $\tau^\eps$ in (5.3) therefore follows by combining 
(5.5), (5.6), and (5.7).

   To derive the lower bound on $\tau^\eps$ in (5.3), first use the
fact that
$$
  \tau^\eps
  = \det\!\big( I + \eps^2 B \big)
  = \sum_{S \subset J^\eps} \det\!\big(\eps^2  B_S \big) \,.
  \tag 5.8
$$
Because $G^\eps$ is Hermitian positive, so is $B =G^{\eps\,2}$.  
Hence, each term of the sum in (5.8) is positive.
We then use the fact that
$$
  B^\eps_S = B^\eps_{SS}
  = \big( G^\eps_{SS} G^\eps_{SS} 
                 + G^\eps_{SS'} G^\eps_{S'S} \big) \,,
$$
and the fact that $G^\eps_{SS}G^\eps_{SS}$ and 
$G^\eps_{SS'}G^\eps_{S'S}$ are Hermitian nonnegative matrices to 
bound each term below by 
$$
\aligned
  \det\!\big(\eps^2 B_S \big) 
  & \geq \det\!\big( \eps^2 G^\eps_{SS} G^\eps_{SS} \big)
\\
  & = \eps^{2 |S|} \big[ \det\!\big( G_S^\eps \big) \big]^2
\\
  & \geq \eps^{2 N^\eps} \big[ \det\!\big( G_S^\eps \big) \big]^2 \,.
\endaligned
  \tag 5.9  
$$
Because each term of the sum in (5.8) is positive, the sum may be
bounded below by the largest lower bound on any of its terms.  By
(5.1) that is achieved for the $S$ where
$$
  \big[ \det\!\big( G_S^\eps \big) \big]^2
  = \exp\!\bigg( \frac{q^\eps(x,\tt)}{\eps} \bigg) \,,
$$
whereby the desired lower bound on $\tau^\eps$ in (5.3) is obtained,
and the proof is complete.

{\bf 5.3. Properties of the Maximization Problem.}

   Before we can characterize the semiclassical limit of the conserved
densities and fluxes, more basic facts concerning the maximization
problem in (5.1) must be established, including the uniqueness of the
density at which the maximum is attained.  By introducing the scalar
product
$$
  (\alpha \,|\, \beta) 
  \equiv \frac1\pi \int_0^1 \alpha(\eta) \, 
                                \beta(\eta) \, d\eta \,,
  \tag 5.10
$$
and the integral operator
$$
  L \psi(\eta) 
  = \frac{1}{2\pi} 
    \int_0^1 
        \log\!\bigg( \frac{\eta -\nu}{\eta + \nu} \bigg)^2 
        \psi(\nu) \, d\nu \,.
  \tag 5.11
$$
The underlying functional being maximized in (5.1) can be recast more 
abstractly as the quadratic form
$$
  Q(\psi;x,\tt) 
  = \tfrac12 (\psi \,|\, \L \psi) + (a(x,\tt) \,|\, \psi) \,.
  \tag 5.12
$$
where $a$ is again given by (4.6) but here we suppress the $\eta$
dependence.  The properties of these objects that we will need are 
presented in the following three lemmas.

   The maximization problem (5.1) in this form  is then posed over the set 
$\AA$ of admissible $L^1$ densities, 
$$
\AA = \{\psi \in L^1[(0,1)]: 0 \leq \psi(\eta)) \leq \phi(\eta)\}, 
$$
which naturally inherits a topology from the weak-$*$ topology of measures.  
However, the densities in $\AA$, being all bounded above by $\rho(\eta)$, are
equi-integrable and, hence, comprise a relatively compact set in the weak
topology of $L^1([0,1])$.  This means that the weak-$*$ topology of
measures and the weak topology of $L^1$ coincide on $\AA$.  We will
always consider $\AA$ equipped with this topology.  The key fact we
will need is stated in the first lemma, the proof of which is omitted.

{\narrower
\smallskip
\noindent
{\bf Lemma 5.2 ($\AA$-compactness).} The set $\AA$ is compact.

\smallskip}

\bigskip

   The operator $\L$ defined by (5.11) and the functional $Q$ defined
by (5.12) are well behaved over the set $\AA$.  Indeed, given that
$\psi$ is in $\AA$, it can be shown [LL2] that $\L\psi$ is in the
class of continuous, odd, functions of $\eta$, which
we denote as $C_{odd}(\R)$.  Moreover, it is
clear from (4.6) that for every $(x,\tt)\in\R\times\TT_o$ the
function $\eta \mapsto a(\eta,x,\tt)$ is also in $C_{odd}(\R)$.
Consequently, $Q$ defined by (5.12) takes values in $\R$.  Moreover,
$\L$ and $Q$ possess continuity properties that are stated in the next
lemma.

{\narrower
\smallskip
\noindent
{\bf Lemma 5.3 ($\AA$-continuity).}

\item{(a)} The operator $\L:\AA\to C_{odd}(\R)$ is continuous.

\item{(b)} The functional $Q:\AA\times\R\times\TT_o\to\R$ 
           is continuous.
  
\smallskip}
\noindent
{\bf Proof.}  The hardest part is the proof of (a) which resembles
that of Theorem 3.4 in [LL2].  The proof of
(b) follows from (a) and the continuity of $a$.  //
\bigskip
\noindent
{\bf Remark.}  By combining Lemma 5.2 with Lemma 5.3, it is seen that
the image of $\AA$ under $\L$ is a compact subset of $C_{odd}(\R)$ and that 
for every $(x,\tt)\in\R\times\TT_o$ the map $\psi \mapsto Q(\psi;x,\tt)$ 
has a compact range and, hence, attains a maximum over $\AA$.  The
existence of the maximum for (5.1), which has already been established
through the limiting procedure of Theorem 5.1, is thereby
re-established intrinsically.  

\smallskip

   The uniqueness of the maximum in (5.1) follows from the third
lemma.

{\narrower
\smallskip
\noindent
{\bf Lemma 5.4 (strict concavity).}

\item{(a)} The quadratic form $(\psi\,|\,\L\psi)$ is negative definite
           over $\AA\ominus\AA$.

\item{(b)} For every $(x,\tt)\in\R\times\TT_o$, the map 
           $\psi\mapsto Q(\psi;x,\tt)$ is strictly concave over $\AA$.
  
\smallskip}
\noindent
{\bf Proof.}  The hardest part is again the proof of (a) which here 
resembles that of Theorem 3.7 in [LL2].  The proof of (b) follows directly 
from (a).  //

\bigskip

   With the above pieces in place, the main result of this subsection
can now be persented.

{\narrower
\smallskip
\noindent
{\bf Theorem 5.5 (uniqueness and regularity).}  For each 
$(x,\tt)\in\R\times\R^\infty$ there exists a unique $\psi^*(x,\tt)$ 
in the admissible class $\AA$ such that
$$
  q(x,\tt) = Q(\psi^*(x,\tt);x,\tt) \,.
  \tag 5.13
$$
Moreover, this maximizing density satisfies the following.

\item{(a)} The map $(x,\tt)\mapsto\psi^*(x,\tt)$ is continuous from 
           $\R\times\TT_o$ into $\AA$ equipped with the weak $L^1$
           topology.

\item{(b)} The map $(x,\tt)\mapsto \L\psi^*(x,\tt)$ is continuous from 
           $\R\times\TT_o$ into $C_{odd}(\S)$ equipped with the 
           uniform topology.

\item{(c)} The map $(x,\tt)\mapsto q(x,\tt)$ is differentiable from 
           $\R\times\TT_o$ into $\R$ with its continuous partial
           derivatives given by

$$
\align
  \del_x q(x,\tt) & = - (\eta \,|\, \psi^*(x,\tt)) \,, 
  \tag 5.14a
\\
  \del_{t_{2n-1}} q(x,\tt) 
  & = (\eta^{2n-1} \,|\, \psi^*(x,\tt)) \,.
  \tag 5.14b
\endalign
$$

\smallskip}
\noindent
{\bf Proof.}  Lemma 5.4 states that for every $(x,\tt)$ the functional
$Q(\cdot;x,\tt)$ is strictly concave over $\AA$, thereby insuring the
uniqueness of the point in $\AA$ at which the maximum in (5.1) is
attained.  Denote this point as $\psi^*(x,\tt)$.  

   To prove the continuity asserted in (a), let $(x,\tt)$ be any point
and $\{(x_k,\tt_k)\}$ any sequence converging to that point in
$\R\times\TT_o$.  Because, by Lemma 5.2, $\AA$ is compact, the
sequence $\{\psi^*(x_k,\tt_k)\}$ has cluster points, all of which lie
in $\AA$.  Let $\psi_*$ denote one such cluster piont.  Upon passing
to a subsequence if necessay, part (b) of Lemma 5.3 implies
$$
  \lim_{k\to\infty} q(x_k,\tt_k) 
  = \lim_{k\to\infty} Q(\psi^*(x_k,\tt_k);x_k,\tt_k)
  = Q(\psi_*;x,\tt) \,.
  \tag 5.15
$$
On the other hand, the continuity of $q$ implies  
$$
  \lim_{k\to\infty} q(x_k,\tt_k) = q(x,\tt) \,.
  \tag 5.16
$$
Comparing (5.15) and (5.16) shows that $Q(\psi_*;x,\tt)=q(x,\tt)$,
whereby we conclude that $\psi_*=\psi^*(x,\tt)$.  Hence, the original 
sequence $\{\psi^*(x_k,\tt_k)\}$, having $\psi^*(x,\tt)$ as the only 
cluster point, must converge to $\psi^*(x,\tt)$.  The continuity
asserted in (a) follows immediately, while that of (b) does so after
invoking the part (a) of Lemma 5.3. 

   Now turn to the differentiability (c).  It suffices to establish 
(5.14b).  Let $\tt$ and $\tt'$ differ only in the coordinate $t_{2n-1}$.
For every $\eta$ in $\AA$, a direct calculation then yields
$$
  Q(\psi;x,\tt') - Q(\psi;x,\tt)
  = (\eta^{2n-1}\,|\, \eta) \, (t_{2n-1}' - t_{2n-1}) \,. 
  \tag 5.17
$$
However, because $\psi^*$ maximizes $Q$, one has the general two-sided
inequality
$$
\aligned
  Q(\psi^*(x,\tt);x,\tt') - Q(\psi^*(x,\tt);x,\tt)
  & \leq q(x,\tt') - q(x,\tt)
\\
  & \leq  Q(\psi^*(x,\tt');x,\tt') - Q(\psi^*(x,\tt');x,\tt) \,,
\endaligned
  \tag 5.18
$$ 
which, when combined with (5.17), gives
$$
\aligned
  \big( \eta^{2n-1} \,\big|\, \psi^*(x,\tt) \big) \, 
  (t_{2n-1}' - t_{2n-1})
  & \leq q(x,\tt') - q(x,\tt)
\\
  & \leq \big( \eta^{2n-1} \,\big|\, \psi^*(x,\tt') \big) \, 
         (t_{2n-1}' - t_{2n-1}) \,.
\endaligned
  \tag 5.19
$$
The result now follow by the continuity of $\eta^*$.  //

\bigskip

   It will prove useful that the solution $\psi^*(x,\tt)$ of the
maximization problem can be characterized in terms of variational
conditions.
  
{\narrower
\smallskip
\noindent
{\bf Theorem 5.6 (variational conditions).}  If $\eta\in\AA$ then
$\eta=\eta^*(x,\tt)$ if and only if $\eta$ satisfies the variational
conditions
$$
  \psi = \cases 0                                          &  
                \text{where $a(\eta,x,\tt)+\L\psi<0$} \,,  \\
                \phi(\eta)                  &
                \text{where $a(\eta,x,\tt)+\L\psi>0$} \,.  \endcases
  \tag 5.20   
$$
\smallskip}
\noindent
{\bf Proof.}  The proof follows that of
Theorem 3.12 in [LL2].  It uses the continuity of $a+\L\psi$ in $\eta$
to assert that the conditionals in (5.20) define open sets. //

\bigskip
\noindent
{\bf 5.4. The Limit of the Densities and Fluxes.}
By combining the differentiability result of Theorem~5.5 with the
following elementary, but nontrivial, lemma will yield a strengthening
in the sense of the convergence for the tau-functions in (5.2).
  
{\narrower
\smallskip
\noindent
{\bf Lemma 5.7 (converging derivatives).}  Let $\{h_n\}$ be a
sequence of differentiable convex functions over $\R^\infty$ such that
$h_n(\xx)\to h(\xx)$ uniformly over compact subsets of
$\xx\in\R^\infty$ and $h$ is differentiable.  Let
$\del_s=\dot\xx\cdot\del_\xx$ denote the directional derivative of
$\xx$ in a given direction $\dot\xx\in\R^\infty$.  Then
$$
  \del_s h_n(\xx) \to \del_s h(\xx) \,,
  \tag 5.21
$$
uniformly over compact subsets of $\xx\in\R^\infty$.

\smallskip}
\noindent
In particular, we can now give the main result of this section.

{\narrower
\smallskip
\noindent
{\bf Theorem 5.8 (limit of densities and fluxes).}
The limit of the tau-function $\tau^\eps$ is given by 
$$
  \lim_{\eps\to 0} \eps^2 \del_{t_{2n-1}} \log\tau^\eps(x,\tt)
  = \del_{t_{2n-1}} q(x,\tt) 
  = \big( \eta^{2n-1} \,\big|\, \psi^*(x,\tt) \big) \,,
  \tag 5.22
$$
uniformly over compact subsets of $(x,\tt)\in\R\times\TT_o$.  The
densities $\rho_n^\eps$ and fluxes $\mu_{m,n}^\eps$ have the
distributional limits
$$
\align
  \text{$\DD'(dx)$--}\lim_{\eps\to 0} \rho_{2n-2}^\eps 
  & = \del_x (\eta^{2n-1} \,|\, \psi^*) \,,
  \tag 5.23a
\\
  \text{$\DD'(dt_{2m-1})$--}\lim_{\eps\to 0} \mu_{2m-1,2n-1}^\eps 
  & = - \del_{t_{2m-1}} (\eta^{2n-1} \,|\, \psi^*) \,.
  \tag 5.23b
\endalign
$$

\smallskip}
\noindent
{\bf Proof.}  Assertion (5.22) follows directly from (5.2) of Theorem
5.1 (tau-function limit) and Theorem 5.14 (uniqueness and regularity)
upon applying Lemma 5.7 (converging derivatives).  Assertion (5.23)
then follows by the usual integration by parts argument.  //

\smallskip
\noindent
{\bf Remark.}  The significance of (5.23) is that one can now pass to 
the limit in the local conservation laws   
$$
  \del_{t_{2m-1}} \rho_{2n-2}^\eps
  + \del_x \mu_{2m-1,2n-1}^\eps = 0 \,, \qquad 
  \text{for $m,n=1,2,\cdots$} \,. 
  \tag 5.24
$$

\bigskip
\bigskip
\centerline{\bf 6. The Limiting Dynamics}
\smallskip

   By the results of the previous section we have seen that  
the limit related to $\tau$ can be converted to a variational problem.
This can in turn be converted to a Riemann-Hilbert boundary value problem
which can be solved with the use of the Hilbert transform. We will list 
some related results but omit most proofs. The interested reader
can find more details in [LL] and [LLV].

  The variational conditions (5.20) for the maximization problem can be 
expressed  
$$
\aligned
   L\psi(\eta) + a(\eta,x,\tt) < 0 \,, \qquad 
    & \text{when $\psi(\eta,x,\tt) = 0 $,}
\\
   L\psi(\eta) + a(\eta,x,\tt) = 0 \,, \qquad 
    & \text{when $0 < \psi(\eta,x,\tt) < \phi(\eta)$,} 
\\
   L\psi(\eta) + a(\eta,x,\tt) > 0 \,, \qquad 
    & \text{when $\psi(\eta,x,\tt) = \phi(\eta)$.}
\endaligned
  \tag 6.1
$$ 
Let $I$ be the interior of the set of $(\eta,x,\tt)$ in which
equality holds and ${\bar I}$ be its closure. Directly differentiating
(6.1) while using formula (4.6) for the function $a$ leads to 
$$
\aligned
  L\psi_x(\eta) = \eta \,, \qquad 
  L\psi_{\tt}(\eta) 
  = - \del_{\tt} \chi(\eta;\tt) \,, \qquad 
   & \text{when $(\eta,x,\tt) \in I$,}
\\ 
  \psi_x(\eta) = 0 \,, \qquad \quad 
   \psi_{\tt}(\eta) = 0 \,, \qquad \qquad \quad \ \,
   & \text{when $(\eta,x,\tt) \notin {\bar I}$,} 
\endaligned
  \tag 6.2
$$ 
where $\chi(\eta; \tt)$ is defined as $\chi_j (\tt)$ if
$\eta = \eta_j$. 

These differentiated variational conditions do not have any explicit
dependence on $(x,\tt)$, but rather their dynamics, as well as
all their memory of the initial data, is contained in the set $I$.
Consequently, these conditions have more general validity than the
variational conditions (6.1). Indeed, the latter vary (although in a
sense insignificantly) when the initial data are considered in
different classes (such as periodic or tending to different limits as
$x \to \pm\infty$), while the differentiated conditions remain
unchanged.

Making the ansatz that at fixed $(x,\tt)$ the set $I(x,\tt)$,
defined as 
$$ 
   I(x,\tt) \equiv \bigg\{ \eta \in [0,1) \,:\,
      (\eta,x,\tt) \in I \bigg\} \,, 
$$ 
consists of a finite union of disjoint open intervals, one can
uniquely determine the functions $\psi_x(\eta)$ and
$\psi_{\tt}(\eta)$ in terms of the endpoints of these
intervals. More precisely, $I(x,\tt)$ is assumed to take the
form
$$
  I(x,\tt) = [0,\beta_1) \cup (\beta_2,\beta_3) \cup \cdots 
       \cup (\beta_{2g},\beta_{2g+1}) \, 
  \tag 6.3
$$
for some nonnegative integer $g$, where the $\beta_i$ and $g$ depend
on $(x,\tt)$ and 
$$
   0 < \beta_1 < \beta_2 < \cdots < \beta_{2g+1} < 1 \,.
$$ 

The operator $L$ given by (5.11) plays an important role for seeking
the unique solution of the maximum problem. 

Extend $\psi(\eta)$ to be an odd function over the real axis $\R$ that
vanishes when $|\eta| \ge 1$, then the $\eta$-derivative of
$L$ is the Hilbert transform:  
$$ 
  \frac{d}{d\eta} L\psi(\eta) 
  = \frac{1}{\pi} 
    \int_{-\infty}^\infty \frac{\psi(\mu)}{\eta - \mu} \, d\mu 
  \equiv H\psi(\eta) \,.  
$$ 
This observation allows the differentiated variational conditions
(6.2) to be transformed into Riemann-Hilbert problems through which
they can be solved explicitly for $\psi_x$ and $\psi_{\tt}$ in
terms of hyperelliptic functions involving a radical function $R$.

   This $R$ has the following form
$$ 
  R(\eta,x,\tt) 
  = \bigg( \prod_{i=1}^{2g+1} 
           \bigg( \beta_i^2 - \eta^2 \bigg) \bigg)^{1/2} \,,
$$ 
where the $\{\beta_i\}$ are the boundaries of the set $I(x,\tt)$ as
given by (6.3). 

Note that $I(x,\tt)$ is the set over which $R(\eta)$ is real valued  
and $g$ is just the genus of the associated Riemann surface.  

The variational problem (5.12) is uniquely solvable, say, by $\psi^{*}$.
It follows from (5.14) and (5.23), for n=1, that all the limits of nontrivial 
conserved densities along odd flows can be represented by the derivatives of 
$\psi^{*}$ in the sense of distributional convergence:
$$
   \rho(x,\tt) = - (\eta | \psi^{*}_{x}),
$$     
$$
   \epsilon(x,\tt) = {2 \over 3}(\eta | \psi^{*}_{t_3}),
$$
$$
   \rho_{2m-2}(x,\tt) = (\eta |  \psi^{*}_{t_{2m-1}}),
$$
for $m = 1, 2, \cdots$.

The integrals above are calculated on $I(x,\tt)$, the
particular domain of $\eta$ as given by (6.3). These $\{\beta_j\}$, 
as the boundary points of $I(x,\tt)$, are then governed by a
genuine nonlinear hyperbolic system under any odd flow. Different flow
makes different time evolution of the domain $I(x,\tt)$ through
its boundary $\{\beta_i\}$, and enforces different time evolution of
the semiclassical limit.

The dependence of the $\beta_i$ on $(x,\tt)$ is then derived from 
the compatibility constraint 
$\del_{\tt}\psi_x=\del_x\psi_{\tt}$. 
This reduces to the first order system of hyperbolic equations in the  
Riemann invariant (diagonal) form
$$
  \del_{t_m} \beta_i 
  + S_{mi}(\beta_1,\cdots,\beta_{2g+1}) \, \del_x \beta_i 
  = 0 \,, \qquad 
  \text{for $i = 1,\cdots,2g+1$} 
  \tag 6.4
$$
for any positive odd number $m$.

In this way, the initial value of $\psi^{*}$ can be represented in the form 
$$
  \psi^*(x,0) 
  = \cases  0 & \text{for $x \geq \Xp(\eta)$} \,, \\
           \int^{\Xp(\eta)}_x \dfrac\eta{(A^2(y) - \eta^2)^{1/2}} \,dy
              & \text{for $\Xp(\eta) \geq x \geq \Xm(\eta)$} \,, \\
           \phi(\eta) & \text{for $x \leq \Xm(\eta)$} \,, 
    \endcases
$$
where $\Xm(\eta) \leq \Xp(\eta)$ are implicitly defined by
$$
   A(\Xp) = A(\Xm) = \eta \, .
$$

Consider the semiclassical limit along the $t_3$ flow, i.e. the mKdV
flow (1.9). Note that its formal limit equation (1.12) has no classical
solution beyond a finite time, say $t=t_b$, because of its hyperbolic
nature. If $\rho(x,\tt)$ is the solution of equation (1.12)
(focusing) with the initial value $\rho(x, 0) = A^2(x)$, then 
$t_b$ is given by
$$ 
   t_b = [\max_{x} (3 AA_x)]^{-1}.
$$

For $t < t_b$, i.e. $g=0$, we have that $\beta^2_1(x,t) =
\rho(x, t)$, an exact solution of (1.2), and
$$
  L^2-\lim_{\eps\to 0} \rho^\eps(x, t) = \beta_1^2(x,t) 
$$
uniformly in $t$.

For $t \geq t_b$, the limits of all conserved densities and their
fluxes exist in the sense of $L^2$ weak convergence.  They are
determined by the $2g + 1$ functions $\{\beta_i\}$ which are governed
by a strict genuinely nonlinear hyperbolic system (6.4) and will break
down again after a finite time.

\bigskip
\bigskip
\noindent
\centerline{{\bf 7. An Invariant Subspace for the Hierarchy}}
\smallskip

   Loop algebras provide a natural ``phase space'' in which to study
the focusing Zakharov-Shabat (ZS) hierarchy.  We will see that, when
$u_{in}$ is restricted to be real, the correspondingly restricted
phase space is invariant under the action of the odd flows of the
hierarchy.  Reduction to this invariant subspace provides a framework
in which to geometrically understand the dispersionless limit of these
odd flows onto the corresponding limit for KdV.

\bigskip
\noindent
{\bf 7.1. Loop Algebra Structure.}  We begin by summarizing the loop
algebra structure for these soliton equations.  Further details may be
found in [FNR].  To facilitate our presentation, we introduce the
following notation for the Pauli basis of $su(2)$:
$$
  {\Cal H} = \pmatrix i & 0 \\ 0 & -i \endpmatrix \,, \qquad 
  {\Cal F} = \pmatrix 0 & 1 \\ -1 & 0 \endpmatrix \,, \qquad
  {\Cal E} = \pmatrix 0 & i \\  i & 0 \endpmatrix \,.
$$
With respect to this basis the ZS spectral problem can be expressed as
$$ 
  \LL - \zeta = - \eps \Cal{H} \del_x + \alpha \Cal{E} 
                                      + \beta \Cal{F} - \zeta \,, 
  \tag 7.1
$$
where $u^\eps(x) = \alpha(x) + i \beta(x)$.  Upon multiplying this 
expression by $\Cal{H}$, we get an equivalent operator,
$$
  \eps \, \del_x - \alpha \Cal{F} + \beta \Cal{E} - \zeta \Cal{H} \,. 
  \tag 7.2
$$
Observe that the nondifferential part of this operator is a real
linear combination of the Pauli matrices, and so is a family of
elements of $su(2)$.  This observation motivates the definition of the
loop algebra for the hierarchy, which is defined as the vector space
of all formal Laurent series, $Q$, with coefficients in $su(2)$:
$$
  Q = \sum_{n= -\infty}^\infty  
          \zeta^n \left( h_n {\Cal H} + a_n {\Cal F} 
                                      + b_n {\Cal E} \right) \,. 
  \tag 7.3
$$
The ``algebra'' structure is a Lie algebra structure induced by the
usual Lie bracket, $[\,\cdot\,,\,\cdot\,]$ for $su(2)$: if 
$Q=\sum_{j=-\infty}^\infty \zeta^j Q_{-j}$, and 
$R=\sum_{k=-\infty}^\infty \zeta^k R_{-k}$, then
$$
  [Q, R] = \sum_{m = -\infty}^\infty 
               \sum_{j+k=m} \zeta^m [Q_{-j}, R_{-k}] \,. 
$$ 
This loop algebra is the natural setting in which to develop the {\it
Lax pair} formulation of the NLS flows.  For example, in the case of
the original focusing NLS flow if we let
$$
  \eps \, \del_x + Q^{(1)}
$$
denote the operator (7.1), then $Q^{(1)}$ is an element of the loop 
algebra (7.3) which is linear in $\zeta$.  There is another operator,
$$
  \eps \, \del_t + Q^{(2)} \,,
$$
where  
$$
\aligned
  Q^{(2)} 
  & = \zeta^2 Q_0 + \zeta Q_1 + Q_2
\\
  & = \zeta^2 {\Cal H} 
      + \zeta \pmatrix 0 & -u  \\ \bar{u} & 0 \endpmatrix
      - \pmatrix \dfrac{i}{2} |u|^2 & 
        i \dfrac{\eps}{2} \del_x u  \\
        i \dfrac{\eps}{2} \del_x \bar{u}  & 
        \dfrac{-i}{2} |u|^2         \endpmatrix 
\endaligned
  \tag 7.4
$$ 
is a quadratic polynomial in the loop algebra (7.3) such that the
compatibility condition,
$$
  \big[ \del_x - Q^{(1)}, \del_t - Q^{(2)} \big] 
  = Q^{(1)}_t - Q^{(2)}_x + [Q^{(1)}, Q^{(2)}] = 0 \,,
$$
known as a Lax equation, is equivalent to the requirement that
$u_\eps$ should solve (1.4a).

\bigskip
\noindent
{\bf 7.2. The Hierarchy of Densities and Flows.} 
In Section 3 we took an approach for constructing the hierarchy of
flows which is based on a Hamiltonian formalism applied to a
generating function for the conserved densities.  However, this
recursive construction can be encoded [FNR] in the loop algebra
representation as follows.  The hierarchy of conserved densities,
$\rho_m = h_{m+2}$ is recursively defined by
$$
  h_n = i \int (u e_n - \bar{u} \bar{e_n}) \, dx \,, 
  \tag 7.5a 
$$
where
$$
  e_n = i/2 (\eps \del_x e_{n-1} - 2i \bar{u} h_{n-1}) \,, 
  \tag 7.5b
$$   
with $h_0 = -1$ and $e_0 = 0$.  Setting $e_n = a_n + i b_n$ then 
determines the coefficients, $a_n, b_n, h_n$ of a loop element (7.3). 

   We now introduce the matrix 
$$
  Q_n = \pmatrix   i h_n    &  e_n   \\ 
                 -\bar{e}_n & -i h_n \endpmatrix \,, 
  \tag 7.6
$$
where $e_n$ and $h_n$ are the quantities we have just defined.  One
sees immediately from the definition of these quantities that $Q_n$ is
an element of $su(2)$; i.e., it is skew-Hermitian.  Using these
recursive definitions it is also a straightforward exercise, which we
leave to the reader, to show that the $Q_n$ satisfy a hierarchy of
coupled ODE's:
$$
  \del_x Q_n - [Q_1, Q_n] = [Q_0, Q_{n+1}] \,. 
  \tag 7.7
$$
This hierarchy can be encoded in a single ``generating'' ODE
$$
  \del_x Q  - [\zeta Q_0 + Q_1, Q] = 0 \,,     
  \tag 7.8
$$
where $Q=Q_0 +\zeta^{-1} Q_1 +\zeta^{-2} Q_2 + \cdots$ is a Laurent
element of our loop algebra, as is again straightforward to check.
The coefficient of $\zeta^{-n}$ on the left-hand side of (7.8) is
equivalent to the $nth$ ODE of (7.7).  Note that the linear polynomial
in the bracket of (7.8), $\zeta Q_0 + Q_1 = Q^{(1)}$ is the
nondifferential part of ${\Cal H} (\LL - \zeta)$ which is equivalent
to the original Zakharov-Shabat eigenvalue problem.  Moreover, if
$\Psi$ be a fundamental solution of the ZS problem, then [FNR] 
$Q=\Psi {\Cal H} \Psi^{-1}$ solves (7.8).  Thus one sees that the loop 
algebra elements have a realization in terms of ``squared
eigenfunctions'' of the ZS eigenvalue problem and, therefore, the
densities $h_n$ and $e_n$ can, in principle, be read off from the
asymptotic expansion in $\zeta$ of the Jost function for the ZS
problem.

\bigskip
\noindent
{\bf 7.3. Invariance of Real Symmetry for the Odd Flows.} 
Before investigating how the loop algebra structure reflects a
restriction to real initial data, we need to introduce one other
ingredient: a natural inner product on the underlying vector space of
loop elements.  For any two loop algebra elements, 
$Q=\sum_{j=-\infty}^\infty \zeta^j Q_{-j}$ and
$R=\sum_{k=-\infty}^\infty \zeta^k R_{-k}$, this symmetric form is 
defined by
$$
  (Q, R) = -\dfrac{1}{2} \, 
           \tr\!\left( \sum_{j+k=0} Q_{-j} R_{-k} \right) \,. 
  \tag 7.9
$$

   Henceforth it will be assumed that $u(x)$ is real-valued; i.e.
that $u(x)=A(x)$ and $S(x)\equiv0$.  Let's use the recursion formulas
(7.5) to calculate the first few densities under the reality
constraint:
$$
\aligned
  e_0 & = 0  \bigg. 
\\
  e_1 & = u h_0 = - u  
\\
  e_2 & = \dfrac{i}{2} \del_x (-u)  
\endaligned 
  \qquad
\aligned 
  h_0 & = -1 \bigg.
\\
  h_1 & = 0
\\ 
  h_2 & = i \int u (-i/2 \del_x u) - u (i/2 \del_x u) \, dx 
        = \dfrac{1}{2} u^2 \,. 
\endaligned 
  \tag 7.10
$$
By using this as a base step, we can prove 

{\narrower
\smallskip
\noindent
{\bf Proposition 7.1.}
If $u \in \R$, then $e_{ev} \in i\R, e_{od} \in \R$ and 
$h_{od} \equiv 0$.  Here the subscripts $ev$ or $od$ respectively 
stand for any subscript that is even or odd.

\smallskip}
\noindent 
{\bf Proof.}  The proposition is proved by induction with the base
step given by (7.10).  Now assume that the proposition has been
established for $e_{2n-2}$ and $h_{2n-2}$ which are thus,
respectively, pure imaginary and real.  Because, by (7.5b) 
$e_{2n-1}=\tfrac{i}{2}\eps\del_x e_{2n-2}+u h_{2n-2}$ it follows that
$e_{2n-1}$ is real.  From this it follows, using (7.5a), that
$h_{2n-1}$ vanishes identically.  Using (7.5b) again, we have 
$e_{2n}=\tfrac{i}{2}\eps\del_x e_{2n-1} \in i\R$ which completes the
induction.

\bigskip

   Now let $Q^{(ev)}$ and $Q^{(od)}$, denote any linear combination of
loop coefficients of the form $Q_{2j}$ and $Q_{2j-1}$ respectively.
(This includes, as a particular instance, loop algebra elements which
are even in $\zeta$ and odd in $\zeta$ respectively.)  Then based on
the preceding proposition one has the following corollary.

{\narrower
\smallskip
\noindent
{\bf Corollary 7.2.} If $u\in\R$, then $Q^{(ev)}\in\R \cdot \Cal{H}
\bigoplus \R \cdot \Cal{E}$ and thus in particular has pure imaginary
components. $Q^{(od)} \in \R \cdot \Cal{F}$ and therefore has real
components.  Moreover, because $\Cal{H}, \Cal{E}, \Cal{F}$ are an
orthonormal basis of $su(2)$ with respect to the inner product
introduced earlier, one always has $Q^{(ev)} \perp Q^{(od)}$ with
respect to this inner product.

\smallskip}
\noindent 
{\bf Proof.} By the proposition, $h_{2j-1}$ is identically zero while
$e_{2j-1}$ is real.  Using this in (7.6) one immediately observes that
$Q_{2j-1}$ is in the span of $\Cal{F}$.  Because the $h$-coefficients
are always real, $h_{2j} \Cal{H}$ is pure imaginary.  Because, by the
proposition, $e_{2j}$ is pure imaginary, it follows that $Q_{2j}$ is
in the span of $\Cal{H}$ and $\Cal{E}$.

\bigskip

   There are equations similar to (7.8) which give the evolution of
the density generating function, $Q$, under the higher flows.  These
have the form [FNR]:
$$
  Q_{t_n} = [Q_{2n-1}, Q] \,. 
  \tag 7.11
$$
The loop $Q$ has a natural decomposition $Q = Q^{(ev)} + Q^{(od)}$
into, respectively, the even and odd terms of the series.  For $\zeta$
real, these are also, respectively, the imaginary and real components
of $Q$.  Moreover, these two parts are orthogonal with respect to the
inner product, $(\cdot,\cdot)$ on the loop algebra.  This decomposition
can be applied to (7.11) (for instance by equating imaginary and real
parts on both sides of the equation).  We apply this in the case of
the odd flows; i.e., when $n = 2m-1$ is odd.  This yields two
equations:
$$
\aligned
  Q^{(ev)}_{t_{2m-1}} & = [Q^{(ev)}, Q_{2m-1}] \,,
\\
  Q^{(od)}_{t_{2m-1}} & = [Q^{(od)}, Q_{2m-1}] = 0 \,. 
\endaligned \tag 7.12
$$
The right-hand side of the last equation is identically zero because 
$[\Cal{F},\Cal{F}]=0$.

   Thus the subspace of even densities is preserved under the odd
flows.  In particular
$$
  Q_2 = \pmatrix \dfrac{i}{2} u^2 & 
                 \dfrac{i}{2} \eps \del_x u \\ 
                 \dfrac{i}{2} \eps \del_x u & 
                 -\dfrac{i}{2} u^2 \endpmatrix \,, 
  \tag 7.13
$$
which carries the densities that comprise the Miura transformation
(2.7), $v=u^2+i\eps\del_x u = 2(h_2 + e_2)$, preserves its form under
the odd flows.  In Section 2 we only considered the evolution of
these densities under the $t_3$-flow, the mKdV flow.

   Notice that for any of the elements in $Q^{(ev)}$, the coefficients
of terms proportional to $\Cal{E}$ are formally $\OO(\eps)$.  Hence,
in the dispersionless limit the odd NLS flows acting on real initial
data collapse onto the Abelian subalgebra spanned by $\Cal{H}$ which
is comprised of the Hamiltonian densities introduced in Section 3.
This extends to the higher odd flows our formal observations about the
dispersionless limit of the mKdV flow which motivated the rigorous
analysis of the zero-dispersion limit carried out in Sections 5 and 6.

\bigskip
\bigskip
\noindent
\centerline{{\bf 8. Concluding Remarks}}
\smallskip

There is an alternative approach to showing that the limits
of the focusing mKdV equation (1.6) are very close to those of the
KdV equation (1.1).   

One is the Miura transform [Mr]. The simple relation
$$
   -u =  M(v) \mathrel{\mathop \equiv}  v^2 \pm \eps v_x
$$
provides a link between the semiclassical limits of the
KdV equation and of the defocusing mKdV equation:
$$
   (u_t + 6u u_x + \eps^2 u_{xxx}) + 
   (2v \pm \eps\del_x)(v_t - 6 v^2 v_x + \eps^2 v_{xxx}) = 0.
$$
This shows that the limit of $v^2$, the mass density, (1.7), of the defocusing
mKdV equation (1.6), converges to $\limt u$, the zero dispersion limit
of the solution to the KdV equation (1.1), in the sense of distributions. 
This relation of semiclassical limits, between the defocusing NLS hierarchy 
and the KdV hierarchy, works at least in the case when their initial values
depend only on the continuous spectrum of the associated ZS Dirac operator
and the KdV \Schrodinger operator $\LL_S$ respectively [JLM2]. 

A similar transform,
$$
   u = - M(iv) = v^2 \mp i\eps v_x
  \tag 8.1
$$ 
gives a relation of solutions between a real valued focusing mKdV
equation and a complex valued KdV equation:
$$
   u_t + 6u u_x + \eps^2 u_{xxx} = 
   (2v \mp i\eps\del_x)(v_t + 6 v^2 v_x + \eps^2 v_{xxx}).
$$
It suggests that the limit of $v^2$, the mass density of the focusing mKdV
equation ,
$$
   v_t + 6 v^2 v_x + \eps^2 v_{xxx} = 0,
$$
should converge to the zero dispersion limit of the KdV equation (1.1) as
$\eps \to 0$ in the distributional sense as we have in fact shown, for a 
reasonable class of initial data, in this paper.

\bigskip
\bigskip
\noindent
{\bf Acknowledgements.}  The authors gratefully acknowledge the support
provided to them:  
N.M.E. from the NSF under grants DMS-9626306 and DMS-0073087;
S.J. from the Air Force under grant AFOSR-90-021; 
C.D.L. from the NSF under grants DMS-8914420 and DMS-9803753. 
C.D.L. and W.D.M. thank the Center for Nonlinear Studies (CNLS) at the
Los Alamos National Laboratory for its support during their visits.  
N.M.E. and C.D.L. thank SISSA for its support during the month of Dec 2000.


\bigskip
\bigskip
\noindent
{\bf References.}
\smallskip

\item{[ELZ]}
N.M. Ercolani, C.D. Levermore, and T. Zhang,
{\it The Behavior of the Weyl Function 
     in the Zero Dispersion KdV Limit},
     Commun. Math. Phys. {\bf 183} (1997), 119--143.

\item{[FFM]} 
H. Flaschka, M.G. Forest, and D.W. McLaughlin, 
{\it Multiphase Averaging and the Inverse Spectral Solutions 
     of the Korteweg-de Vries Equation}, 
     Comm. Pure Appl. Math. {\bf 33} (1980), 739--784.

\item{[FNR]} 
H. Flaschka, A.C. Newell, and T. Ratiu, 
{\it Kac-Moody Lie Algebras and Soliton Equations II: Lax 
     Equations Associated with $A_1^{(1)}$}, 
     Physica D {\bf 9} (1983), 300--323.

\item{[FL]} 
M.G. Forest and J.-E. Lee, 
{\it Geometry and Modulation Theory for the Periodic Nonlinear
     \Schrodinger Equation}, 
     in {\it Oscillation Theory, Computation, and Methods of 
             Compensated Compactness}, 
             C. Defermos, J.L. Erickson, D. Kinderleher, and
             M. Slemrod, eds., 
             IMA Volumes on Mathematics and Its Applications {\bf 3},
             Springer-Verlag, New York, 1986, 35--69.

\item{[GGKM]} 
C.S. Gardner, J.M. Greene, M.D. Kruskal, and R.M. Miura,
{\it Method for Solving the Korteweg-deVries Equation},
     Phys. Rev. Lett. {\bf 19} (1967), 1095--1097.

\item{[JLM1]} 
S. Jin, C.D. Levermore, and D.W. McLaughlin,
{\it The Behavior of Solutions of the NLS Equation 
     in the Semiclassical Limit},
     in {\it Singular Limits of Dispersive Waves},
             N.M. Ercolani, I.R. Gabitov, C.D. Levermore, 
             and D. Serre eds., 
             NATO ASI Series B {\bf 320}, Plenum, New York, 1994,
             235--255.

\item{[JLM2]} 
S. Jin, C.D. Levermore, and D.W. McLaughlin,
{\it The Semiclassical Limit for the Defocusing Nonlinear
     \Schrodinger Hierarchy}, 
     Commun. Pure \& Appl. Math. {\bf 52} (1999), 613--654.

\item{[KMM]}
S. Kamvissis, K. T.-R. McLaughlin, and P.D. Miller,
{\it Analysis of an Unstable Semiclassical Soliton Ensemble for the
     Focusing Nonlinear \Schrodinger Equation}, 
     Anals of Mathematics Studies, 
     Princeton University Press (to appear).
     
\item{[KS]}
M. Klaus and J.K. Shaw,
{\it Purely Imaginary Eigenvalues of Zakharov-Shabat Systems},
     Phys. Rev. E {\bf 65} (2002).

\item{[LL1]} 
P.D. Lax and C.D. Levermore, 
{\it The Zero Dispersion Limit of the Korteweg-deVries Equation}, 
     Proc. Nat. Acad. Sci. USA {\bf 76} (1979), 3602--3606.

\item{[LL2]} 
P.D. Lax and C.D. Levermore, 
{\it The Small Dispersion Limit of the Korteweg-deVries Equation 
     I, II, III}, 
     Commun. Pure \& Appl. Math. {\bf 36} (1983), 
     253--290, 571--593, 809--829.

\item{[LLV]} 
P.D. Lax, C.D. Levermore, and S. Venakides, 
{\it The Generation and Propagation of Oscillations in Dispersive IVPs 
     and Their Limiting Behavior},
     in {\it Important Developments in Soliton Theory}, 
             T. Fokas and V.E. Zakharov eds., 
             Springer-Verlag, Berlin, 1993, 205--241.

\item{[Lev]}
C.D. Levermore,
{\it The KdV Zero-Dispersion Limit and Densities 
     of Dirichlet Spectra},
     in {\it Recent Advances in Partial Differential Equations 
             and Applications} (a conference celebrating the $70^{th}$
             birthdays of Peter D. Lax and Louis Nirenberg, Venice,
             Italy, 10--14 June 1996), 
             Proceedings of Symposia in Applied Mathematics {\bf 54},
             R. Spigler and S. Venakides eds., 
             American Mathematical Society, Providence, 1998, 187--210.

\item{[Mr]} 
R.M. Miura, 
{\it Korteweg-de Vries equation and generalizations. I. A remarkable 
     explicit nonlinear transformation}, 
     J. Mathematical Phys. {\bf 9} (1968), 1202--1204.

\item{[V1]} 
S. Venakides, 
{\it The Generation of Modulated Wavetrains in the Solution 
     of the Korteweg-deVries Equation}, 
     Commun. Pure \& Appl. Math. {\bf 38} (1985), 883--909.

\item{[V2]} 
S. Venakides, 
{\it The Zero Dispersion Limit of the Korteweg-deVries Equation with
     Periodic Initial Data},  
     AMS Trans. {\bf 301} (1987), 189--225.

\item{[V3]} 
S. Venakides, 
{\it Higher Order Lax-Levermore Theory}, 
     Commun. Pure \& Appl. Math. {\bf 43} (1990), 335--362.

\item{[W]} 
G.B. Whitham, 
{\it Nonlinear Dispersive Waves}, 
     Proc. Royal Soc. London Ser. A {\bf 283} (1965), 238--261.

\item{[ZS1]} 
V.E. Zakharov and A.B. Shabat, 
{\it Exact Theory of Two-dimensional Self-focusing and One-dimensional 
     Self-modulation of Waves in Nonlinear Media}, 
     Sov. Phys. JETP {\bf 34} (1972), 62--69.

\item{[ZS2]} 
V.E. Zakharov and A.B. Shabat, 
{\it Interaction Between Solitons in a Stable Medium}, 
     Sov. Phys. JETP {\bf 37} (1973), 823--828.



\bye
\end